\documentclass[12pt]{article}
\usepackage{color}
\usepackage{hyperref}
\usepackage{amsmath,amssymb,accents}
\definecolor{noir}{gray}{0.} 
\definecolor{myblue}{rgb}{0.14,0.11,0.49}
\definecolor{myred}{rgb}{0.74,0.22,0.15}
\definecolor{mygreen}{rgb}{0.05,0.52,0.42}
\definecolor{myyellow}{rgb}{0.96,0.92,0.13}
\definecolor{myorange}{rgb}{1,0.61,0.36}
\definecolor{mypurple}{rgb}{0.71,0.02,1}
\definecolor{htc}{rgb}{1,1,1} 

\newcommand{\Couleur}[1]{\textcolor{noir}{#1}}
\newcommand{\Mat}[1]{{{\boldsymbol{#1}}}}
\newcommand{\abs}[1]{\left\vert#1\right\vert}
\newcommand{\interior}[1]{\accentset{\smash{\raisebox{-0.12ex}{$\scriptstyle\circ$}}}{#1}\rule{0pt}{2.3ex}}

\def\be{\begin{equation}}
\def\ee{\end{equation}}
\def\bea{\begin{eqnarray}}
\def\eea{\end{eqnarray}}
\def\bc{\begin{center}}
\def\ec{\end{center}}
\def\bi{\begin{itemize}}
\def\ei{\end{itemize}}
\def\bs{\begin{frame}}

\def\dd{\operatorname{d}}

\def\noi{\noindent}
\begin{document}
%
%
\title{On the definition of energy for a continuum, its conservation laws, and the energy-momentum tensor}
\author{
Mayeul Arminjon\\
\small\it Laboratory ``Soils, Solids, Structures, Risks'', 3SR \\
\small\it (Grenoble-Alpes University \& CNRS), Grenoble, France.
} 
\date{}

\maketitle
\begin{abstract}
\noindent We review the energy concept in the case of a continuum or a system of fields. First, we analyze the emergence of a true local conservation equation for the energy of a continuous medium, taking the example of an isentropic continuum in Newtonian gravity. Next, we consider a continuum or a system of fields in special relativity: we recall that the conservation of the energy-momentum tensor contains two local conservation equations of the same kind as before. We show that both of these equations depend on the reference frame, and that, however, they can be given a rigorous meaning. Then we review the definitions of the canonical and Hilbert energy-momentum tensors from a Lagrangian through the principle of stationary action in a general spacetime. Using relatively elementary mathematics, we prove precise results regarding the definition of the Hilbert tensor field, its uniqueness, and its tensoriality. We recall the meaning of its covariant conservation equation. We end with a proof of uniqueness of the energy density and flux, when both depend polynomially of the fields.\\

\noi {\it Keywords:} energy conservation; conservation equation; special relativity; general relativity; Hilbert tensor; variational principle

\end{abstract}

\section{Introduction and summary}

The subject of this paper is wide and there is a huge literature about it. The aim of the paper is to give a unified exposition of what, in this author's view, are the main aspects of the subject, in a relatively short space, while, nevertheless, emphasizing or precising some not widely appreciated facts and providing strict proofs of some less obvious matters (mainly the Hilbert energy-momentum tensor), using not too sophisticated mathematics.\\ 

In non-relativistic classical physics, the concept of energy emerges when one considers the power done (the scalar product of the force by the velocity) on a mass point or a volume element. First, in the schematic case of a mass point in a time-independent potential force field \Couleur{$V$}, the energy of the mass point appears from the power equation as a natural conserved scalar quantity: the sum \Couleur{$\frac{1}{2}mv^2+V$}. That quantity is still relevant if the potential depends on time, but it is not constant any more. This is well known. In the more realistic case of a continuous medium subjected to internal forces and to an external force field, the energy is a volume density and it still emerges from the power done. However, in general, the local conservation of energy then appears in the form of a balance equation, though it is one in which there is no source term. That is, energy conservation means that the energy leaving or entering a given domain is exactly identified as a flux going through the boundary surface of the domain. This also is well known --- see, e.g., Ref. \cite{Mueller&Mueller2009}. We illustrate the emergence of such a true conservation equation for a continuous medium in {\bf Section \ref{NG}} by examining in detail the example of a self-gravitating system of deformable media with isentropic deformation in Newtonian gravity.\\

In relativistic theories (including relativistic quantum mechanics), on the other hand, the volume energy density is essentially the \Couleur{$(0\ 0)$} component of the energy-momentum-stress tensor, in what follows ``the T-tensor" for brevity. The conservation-type equations verified by the T-tensor are discussed in nearly all textbooks about special or general relativity, of course. In {\bf Section \ref{T-Minkowski}}, we recall why the conservation equation verified by the T-tensor in the Minkowski spacetime (see e.g. Landau \& Lifshitz \cite{L&L} or Fock \cite{Fock1959}) contains two true local conservation equations of the form found in the non-relativistic example of Section \ref{NG}\,; and why one may identify the density and flux in these two equations as those of energy and spatial momentum, respectively. We also note the dependence of the energy density and the energy flux on the reference frame. That dependence is a known fact (although a scarcely mentioned one), but often that fact is not well appreciated. Our new contribution here is to show that these quantities can nevertheless be given a rigorous meaning within a theory of general reference frames and the associated space manifolds.\\

As is well known, an expression of the T-tensor may be deduced when a Lagrangian is available, the latter being assumed to govern the relevant system of matter fields via the principle of stationary action (e.g. \cite{L&L,Fock1959, Belinfante1939,Trautman1956,Wald1984,GotayMarsden1992,ForgerRoemer2004,Leclerc2006b}). There are two distinct definitions of a T-tensor from a Lagrangian: (i) The so-called ``canonical" or ``Noether" tensor, say \Couleur{$\Mat{\tau}$}, is a by-product of the Euler-Lagrange equations. (ii) The ``Hilbert tensor", say \Couleur{$\Mat{T}$}, is the symmetric tensor obtained as the derivative of the Lagrangian density with respect to variations of the (spacetime) metric. In {\bf Section \ref{T-tensor}} we review the definitions of the canonical and Hilbert tensors from a Lagrangian through the principle of stationary action in a general spacetime. We recall two important but seemingly not widely known cases where the ``canonical tensor" is, or is not, a tensor. Then we prove precise results regarding the definition of the Hilbert tensor field (\hyperref[Theorem 1]{Theorem 1}). In doing so, we formulate sufficient conditions of regularity for the bounded set in which the action is calculated; we define exact boundary conditions to be verified by the infinitesimal coordinate change; and we give a detailed derivation of the equations. We do not need to use complex notions of fibre bundles. To our knowledge, such a relatively elementary but detailed proof is not available in the literature. Next, we recall the meaning of the standard conservation equation verified by the Hilbert tensor: we argue that one actually needs {\it local} definitions of the energy and momentum densities and their fluxes, in short a local definition of the T-tensor, and one needs also a {\it local} conservation equation for the energy. We briefly discuss a recent work that proposes a solution to the latter issue. We end Section \ref{T-tensor} by stating and proving precise results regarding the uniqueness and the actual tensoriality of the Hilbert tensor (\hyperref[Theorem 2]{Theorem 2}). In particular, we prove that the same variational equation applies when a {\it complete} variation of the {\it sole} metric is applied: Eq. (\ref{delta S special-2}), as when the variation of the metric results from a mere {\it coordinate change} (or diffeomorphism): Eq. (\ref{delta_S_coordinate-change-T}) -- although the meaning of these two equations is totally different, e.g. the l.h.s. of (\ref{delta_S_coordinate-change-T}) is zero for an invariant Lagrangian. We {\it prove} that the variational equation (\ref{delta S special-2}) characterizes the components of the Hilbert tensor field --- whence follows that it is left unchanged by the addition of a four-divergence. We also prove, in detail and by relatively elementary arguments, that this is indeed a \Couleur{$(0\ 2)$} tensor. This is just stated in the literature that we consulted, except for Ref. \cite{GotayMarsden1992} which uses more advanced mathematics.  Of course it follows basically from the invariance of the action but, in our opinion, not in a fully trivial way. \\

Finally, in {\bf Section \ref{Uniqueness E}} we investigate if the energy equation is unique for a given system of fields, i.e., if the energy density and fluxes can be considered to be uniquely defined. We show that, if the energy density and its flux depend on the fields (both the matter fields and the ``long-distance" fields) in a polynomial way, then they are determined uniquely. We show this by considering separately the contributions of matter (including its potential energy in the long-distance fields) and the long-distance fields.

\section{Local energy conservation for an isentropically deformable medium in Newtonian gravity}\label{NG}
\subsection{Local energy balance for the matter fields}

Let us consider a deformable continuous medium, of mass density field \Couleur{$\rho $}, having a general motion (including deformation and rotation), with velocity field \Couleur{${\bf v}$}, with respect to some inertial frame \Couleur{$\mathrm{F}$}. I.e., \Couleur{${\bf v}:=\frac{\dd {\bf x}}{\dd t}$}, where \Couleur{${\bf x}(X):=(x^i)_{i=1,2,3}$} is the spatial position associated with an event \Couleur{$X$} in the frame \Couleur{$\mathrm{F}$}, and where \Couleur{$t \mapsto X(t)$} is the world line of a given ``particle" of the medium, parameterized by the Newtonian time  \Couleur{$t$}. The internal force field in that medium is assumed to be described by the Cauchy stress tensor field \Couleur{$\Mat{\sigma }$}. We assume that this motion takes place in a gravitational field, with Newtonian gravity potential \Couleur{$U$}. Newton's second law for a volume element of the medium writes then:
\be\label{Newton continuum}
\Couleur{\rho\, \frac{\dd {\bf v}}{\dd t}= \rho \,\nabla U  + {\bf div}\, \Mat{\sigma }},
\ee
where $\Couleur{\frac{\dd}{\dd t}}$ means the ``material" (or ``total") derivative: for a vector,
\footnote{\
Of course, \Couleur{$\nabla U$} is the spatial vector having components \Couleur{$(\nabla U)^i=U_{,i} \ (i=1,2,3)$} in any Cartesian coordinate system (``Cartesian components") --- i.e., in any coordinate system for which we have the identity \Couleur{$h_{ij}\equiv \delta _{ij}$}, where \Couleur{$\Mat{h}$} is the (Euclidean) spatial metric. Also, $\Couleur{{\bf div}\, \Mat{\sigma }}$ is the spatial vector with Cartesian components $\Couleur{({\bf div}\, \Mat{\sigma })^i=\sigma ^{ij}_{\ \,,j}}$. And \Couleur{${\bf grad \,v}$} is the mixed spatial tensor having Cartesian components \Couleur{$({\bf grad \,v})^i_{\ \,j}=v^i_{\ \,,j}$}. Mixed tensors are identified with linear mappings: \Couleur{$({\bf grad \,v})({\bf w})$} is the spatial vector having components \Couleur{$({\bf grad \,v})^i_{\ \,j}\,w^j$}, which is the same vector independently of the coordinate system. We use the spatial metric \Couleur{$\Mat{h}$} to raise or lower the indices, e.g. \Couleur{$v_i := h_{ij}v^j$}. The equations in this section are valid in any spatial coordinates, whether Cartesian or curvilinear. 
} 
\be
\Couleur{\frac{\dd {\bf v}}{\dd t}=\frac{\partial  {\bf v}}{\partial  t}+({\bf grad \,v})({\bf v}}). 
\ee
The power (per unit volume) is got by taking the scalar product of (\ref{Newton continuum}) with the velocity \Couleur{${\bf v}$}. On the new l.h.s., we have \Couleur{$\Mat{h}({\bf v},\frac{\dd {\bf v}}{\dd t})=  \frac{\dd }{\dd t}(\frac{{\bf v}^2}{2})$}, while on the r.h.s. we note that
\be
\Couleur{{\bf v.div}\, \Mat{\sigma }:= \Mat{h}({\bf v},{\bf div}\, \Mat{\sigma }) =\mathrm{div}\,\Mat{\sigma }({\bf v})-\Mat{\sigma }{\bf :D}},
\ee
where 
\be
\Couleur{{\bf D} := \frac{1}{2}\left({\bf grad \,v}+({\bf grad \,v})^T \right)}
\ee
is the strain rate tensor.
\footnote{\
\Couleur{$\Mat{\sigma }({\bf v})$} is the vector transformed of \Couleur{${\bf v}$} by the linear mapping associated with the mixed tensor form of \Couleur{$\Mat{\sigma }$}, thus \Couleur{$[\Mat{\sigma }({\bf v})]^i:=\sigma ^i_{\ \, j} v^j=\sigma ^{ij}v_j\,$}. And \Couleur{$\,\Mat{\sigma }{\bf :D}\,:= \sigma^i_{\ \,j}\,D^j_{\ \,i}$} is the scalar product of mixed tensors, defined by double contraction. 
}
We note also that \Couleur{${\bf v.}\nabla U=\frac{\dd U}{\dd t}-\frac{\partial U}{\partial t}$}. Still, we assume that the deformation of the continuum is isentropic, which means that the power of the internal forces is stored as the rate of elastic energy: 
\be\label{Isentropy}
\Couleur{\Mat{\sigma }{\bf :D} = \rho \frac{\dd \Pi }{\dd t}},
\ee 
with \Couleur{$\Pi $} the mass density of internal (elastic) energy in the continuous medium. This assumption applies, in particular, to any {\it elastic medium,} and also \cite{Fock1959} to any {\it barotropic perfect fluid,} as is commonly assumed in astrophysics --- although a barotropic fluid is not plainly an elastic medium in the sense that it does not have a reference configuration. We thus get:
\be
\Couleur{\rho \frac{\dd }{\dd t}\left(\frac{{\bf v}^2}{2}\right)=\rho \left (\frac{\dd U}{\dd t}-\frac{\partial U}{\partial t} \right) +\mathrm{div}\,\Mat{\sigma }({\bf v}) - \rho \frac{\dd \Pi }{\dd t}}.
\ee
It suggests itself to put together the terms containing an exact total derivative:
\be\label{rho de_m/dt}
 \Couleur{\rho \frac{\dd }{\dd t}\left(\frac{{\bf v}^2}{2}+\Pi-U \right)=-\rho \frac{\partial U}{\partial t}  +\mathrm{div}\, \Mat{\sigma }({\bf v})}.
\ee
On the r.h.s., we have a source term due to the external force field, plus a flux term. On the l.h.s., we have 
\be\label{Def e_m}
\Couleur{e_\mathrm{m} := \frac{{\bf v}^2}{2}+\Pi-U}.
\ee
Using the continuity equation that expresses the mass conservation: 
\be\label{Continuity Eqn}
\Couleur{\frac{\partial \rho }{\partial  t} +\operatorname{div} \,(\rho {\bf v})   =0},
\ee
we get easily the well known fact that (for whatever scalar function \Couleur{$e_\mathrm{m}$}, actually):
\be\label{Particular -> Balance}
\Couleur{\rho \frac{\dd e_m}{\dd t} := \rho \left( \frac{\partial e_m }{\partial  t} +{\bf v.}\nabla e_m\right ) = \frac{ \partial (\rho e_m) }{\partial  t} +\operatorname{div} \,(\rho e_m{\bf v}) } .
\ee
From (\ref{Def e_m}) and (\ref{Particular -> Balance}), we may rewrite (\ref{rho de_m/dt}) as
\be\label{Newton Matter Energy Balance}
\Couleur{\frac{\partial w_{\operatorname{m} }}{\partial  t} +\operatorname{div} \Mat{\Phi }_{\operatorname{m}}   =-\rho \frac{\partial U}{\partial t}},
\ee
with 
\be\label{w_m NG}
\Couleur{w_{\operatorname{m}}:= \rho e_m=\rho  \left( \frac{ {\bf v}^2}{2}+\Pi -U \right )}
\ee
and
\be\label{Phi_m NG}
\Couleur{\Mat{\Phi }_{\operatorname{m}}:= w_{\operatorname{m}} {\bf v}-\Mat{\sigma}({\bf v})}.
\ee
That is, we got a balance equation with an external source term on the r.h.s. The scalar field \Couleur{$w_{\operatorname{m}}$} is thus the {\it volume energy density of matter,} including its potential energy in the gravitational field, and the spatial vector field \Couleur{$\Mat{\Phi }_{\operatorname{m}}$} is thus (the surface density of) the {\it matter energy flux}. Equation (\ref{Newton Matter Energy Balance}) can be found in the literature: see Eq. (66.11) in Fock \cite{Fock1959}. But its detailed derivation illustrates well the emergence of a balance equation for a continuous medium.



\subsection{Balance for the gravitational field and local energy conservation equation}

Now we assume that all of the matter that produces the gravitational field is indeed in the form of isentropically deformable continuous media. (Of course, the characteristics of the media may vary in space.) Thus the point-dependent mass density \Couleur{$\rho $} is just the source of the gravitational field.  It therefore obeys the gravitational field equation, i.e. the Poisson equation:
\be\label{Poisson}
\Couleur{\Delta U=-4 \pi G \rho }. 
\ee
By using Cartesian coordinates, for which we have \Couleur{$\Delta U=U_{,i,i}$}, one checks easily that Eq. (\ref{Poisson}) implies the following:
\be\label{Newton Grav Energy Balance}
\Couleur{\frac{\partial w_{\operatorname{g} }}{\partial  t} +\operatorname{div} \,\Mat{\Phi }_{\operatorname{g}}=\rho\, \frac{\partial U}{\partial t}},
\ee
where
\be\label{Def w_g}
\Couleur{w_{\operatorname{g}}:= \frac{  \left( \nabla U \right )^2}{8\pi G}}
\ee 
is the {\it volume energy density of the gravitational field,} and 
\be\label{Def Phi_g}
\Couleur{\Mat{\Phi }_{\operatorname{g}} := -\frac{\partial U}{\partial t}\frac{\nabla U }{4\pi G}} 
\ee
is the {\it gravitational energy flux}. Equation (\ref{Newton Grav Energy Balance}) may be termed the energy balance equation of the gravitational field. Like (\ref{Newton Matter Energy Balance}), this also is a balance equation with a source term. The source term in Eq. (\ref{Newton Grav Energy Balance}) is just the opposite of the source term in (\ref{Newton Matter Energy Balance}). Therefore, combining (\ref{Newton Matter Energy Balance}) with (\ref{Newton Grav Energy Balance}), we get the {\it local energy conservation equation in Newtonian gravity} \cite{A15}:
\be\label{Local Conserv E}
 \Couleur{\frac{\partial w}{\partial t}+\mathrm{div}\,\Mat{\Phi}=0 },
\ee
with the total energy density \Couleur{$w:=w_{\operatorname{m}} + w_{\operatorname{g}}$} and the total energy flux \Couleur{$\Mat{\Phi }:=\Mat{\Phi }_{\operatorname{m}} + \Mat{\Phi }_{\operatorname{g}}$}. Equation (\ref{Local Conserv E}) is the standard form for a true local conservation of energy in a continuum, the definition of the field variables $w$ and $\Mat{\Phi}$ depending on the particular theory. It has essentially the same form as the continuity equation (\ref{Continuity Eqn}). There is also a local conservation equation for momentum in Newtonian gravity, and global (integral) conservation laws can also be derived; see e.g. Refs. \cite{A15,Chandra1969}. Strangely enough, however, we did not see in the literature the local Eq. (\ref{Local Conserv E}) for an elastic medium or a barotropic fluid in Newtonian gravity [thus with the definitions (\ref{w_m NG})--(\ref{Phi_m NG}) and (\ref{Def w_g})--(\ref{Def Phi_g})]. For instance, it is not there in the references quoted in the present paper. (The \Couleur{$i=0$} component of Eq. (13) in Ref. \cite{Chandra1969} is just the continuity equation (\ref{Continuity Eqn}), with \Couleur{$\rho$} being indeed the (Newtonian) density of mass; thus, it is not the conservation equation for the Newtonian energy, but the one for the mass.)


\section{Local conservation equations and the energy-momentum tensor in Minkowski spacetime} \label{T-Minkowski}

\noi Recall that the energy-momentum tensor of a continuum or a system of fields is a second-order spacetime tensor field \Couleur{$\Mat{T}$}, preferably symmetric. In the Minkowski spacetime, \Couleur{$\Mat{T}$} verifies  \cite{L&L} the {\it local conservation equation}
\be\label{True_Cons}
\Couleur{T ^{\mu \nu }_{\ \ ,\nu }=0}\quad  (\operatorname{in\ Cartesian\ coordinates}).
\ee
[Here, Cartesian coordinates are now ones such that the {\it spacetime} metric has components \Couleur{$g_{\mu \nu}=\eta _{\mu \nu  }$}, where the matrix \Couleur{$(\eta _{\mu \nu  }):= \mathrm{diag} (1,-1,-1,-1)$}.] It is easy to see that (\ref{True_Cons}) is the conjunction of two conservation equations having the standard form (\ref{Local Conserv E}). One is precisely the scalar conservation equation (\ref{Local Conserv E}):
\be\label{Local Conserv E-2}
 \Couleur{\frac{\partial w}{\partial t}+\mathrm{div}\,\Mat{\Phi}=0 \quad (t:=\frac{x^0}{c})},
\ee
in which now (the tensor \Couleur{$\Mat{T}$} being taken in mass units as in Fock \cite{Fock1959})
\be\label{Def w Phi}
\Couleur{\ \, w:= c^2 T^{0 0},\quad \Mat{\Phi}:= c^3 T ^{0 i }\partial _i} \quad (\mathrm{sum\ over\ }\Couleur{i=1,2,3)}.
\ee
The other conservation equation involved in (\ref{True_Cons}) and having the form (\ref{Local Conserv E}) is a (spatial) vector equation:
\be\label{Local Conserv P}
\Couleur{\frac{\partial  \Mat{P}}{\partial t} + {\bf div}\,\Mat{\Sigma }=0},
\ee
where 
\be\label{Def P Sigma}
\Couleur{\Mat{P} := c T^{i 0}\partial _i,\quad \Mat{\Sigma }:= c^2 T ^{i j }\partial _i \otimes \partial _j}.
\ee 
We may integrate either of the two conservation equations (\ref{Local Conserv E-2}) and (\ref{Local Conserv P}) in {\it any} bounded spatial domain \Couleur{$\Omega $} (the integrability in an unbounded domain being not guaranteed). This gives us two integral conservation equations:
\be\label{Conserv E}
\Couleur{\frac{\dd }{\dd t} \left(\int_\Omega w\, \dd V \right) =-\int_{\partial \Omega } \Mat{\Phi.n }\, \dd S},
\ee
and
\be\label{Conserv P}
\Couleur{\frac{\dd }{\dd t} \left(\int_\Omega \Mat{P}\, \dd V \right) =-\int_{\partial \Omega } \Mat{\Sigma .n }\, \dd S}.
\ee
Thus in Eqs. (\ref{Conserv E}) and (\ref{Conserv P}), the change on the l.h.s. is due to the flux through the boundary \Couleur{$\partial \Omega $} on the r.h.s.. The scalar \Couleur{$w$} is interpreted as the volume density of energy, and the spatial vector \Couleur{$\Mat{P}$} is interpreted as the volume density of momentum \cite{L&L}. Therefore, in view of (\ref{Conserv E}) and (\ref{Conserv P}), the spatial vector \Couleur{$\Mat{\Phi}$} is interpreted as the surface density of the energy flux, and the spatial tensor \Couleur{$\Mat{\Sigma}$} is interpreted as the surface density of the momentum flux. This interpretation may be justified in several ways, notably the following two \cite{Fock1959}: \\

--- First, by examining the non-relativistic limit for a barotropic perfect fluid or an elastic solid. In the second approximation, \Couleur{$w,\ \Mat{\Phi},\ \Mat{P}$}, and \Couleur{$\Mat{\Sigma}$} in Eqs. (\ref{Def w Phi}) and (\ref{Def P Sigma}) have then the following expressions (\cite{Fock1959}, Sect. 32):
\be
\Couleur{w := c^2\, T^{00} \simeq   c^2\,\rho +\frac{1}{2}\rho {\bf v}^2 +\rho \Pi },
\ee
\be
\Couleur{\Mat{\Phi}:=c^3 T ^{0 i }\partial _i= c^3 T^{i 0}\partial _i = c^2 \Mat{P}   \simeq  c^2 \rho {\bf v}+\left(\frac{1}{2}\rho {\bf v}^2 +\rho \Pi \right){\bf v} - \Mat{\sigma }({\bf v}) },
\ee
\be
\Couleur{\Mat{\Sigma} := c^2 T ^{i j }\partial _i \otimes \partial _j \simeq \rho{\bf v}\otimes {\bf v}-\Mat{\sigma}}.
\ee
Therefore, at the first approximation, the special-relativistic local energy conservation (\ref{Local Conserv E-2}) reduces to the continuity equation (\ref{Continuity Eqn}), and at the second approximation it expresses the conservation of that rest-mass energy corrected by adding the conserved Newtonian energy [Eq. (\ref{Local Conserv E}) with \Couleur{$U\equiv 0$}]. Also, at the first approximation, the special-relativistic momentum conservation (\ref{Local Conserv P}) reduces to the Newtonian momentum conservation equation in the absence of external field:
\be\label{Local Conserv P-Newt-Free}
\Couleur{\frac{\partial \left (\rho {\bf v} \right )}{\partial t} + {\bf div}\,\left( \rho{\bf v}\otimes {\bf v}-\Mat{\sigma} \right )=0}.
\ee

--- Second, by recognizing in \Couleur{$w$} and \Couleur{$\Mat{P}$}, for the electromagnetic field, the usual definition of the electromagnetic energy density and the Poynting vector from the relevant expression of $\Mat{T}$  (\cite{Fock1959}, Sect. 33). \\

One does not use the symmetry of the tensor \Couleur{$\Mat{T}$} to derive Eqs. (\ref{Local Conserv E-2})--(\ref{Local Conserv P}) and (\ref{Conserv E})--(\ref{Conserv P}). If that symmetry is true, it implies that \Couleur{$\Mat{\Phi}=c^2\Mat{P }$} is true generally: the density of energy flux is equal to \Couleur{$c^2$} times the density of momentum. The same Eqs. (\ref{Conserv E}) and (\ref{Conserv P}) apply also to Newtonian gravity, as follows from Eqs. (\ref{Local Conserv E}) and the gravitational extension \cite{A15, Chandra1969} of (\ref{Local Conserv P-Newt-Free}).\\

\vspace{2mm}
As is easy to check, under a purely spatial change of the chart (coordinate system):
\be\label{purely-spatial-change}
\Couleur{x'^j=f^j((x^k))\ (j,k=1,2,3)}\ \mathrm{(or} \ \Couleur{{\bf x}'={\bf f}({\bf x})}),\qquad \mathrm{and}\quad \Couleur{x'^0=x^0}, 
\ee
the energy density \Couleur{$w$} [Eq. (\ref{Def w Phi})$_1$] is an invariant scalar, while \Couleur{$\Mat{\Phi }$} [Eq. (\ref{Def w Phi})$_2$] and \Couleur{$\Mat{P}$} [Eq. (\ref{Def P Sigma})$_1$] transform indeed as spatial vectors; and \Couleur{$\Mat{\Sigma }$} [Eq. (\ref{Def P Sigma})$_2$] transforms indeed as a \Couleur{$(2\ \,0)$} spatial tensor. One may give a rigorous geometric meaning to such ``spatial" objects by defining a relevant {\it space manifold} $\mathrm{M}_\mathrm{F}$, as follows \cite{A44}. In a {\it general} spacetime, one can formally define a {\it reference frame} \Couleur{$\mathrm{F}$} as being an equivalence class of charts having the same domain of definition \Couleur{$\mathrm{U}$} (an open subset of the spacetime manifold \Couleur{$\mathrm{V}$}) and exchanging by a coordinate change (``transition map") having the form (\ref{purely-spatial-change}). Let $P_S: \mathbb{R}^4 \rightarrow \mathbb{R}^3, {\bf X}:= (x^\mu )\mapsto {\bf x}:= (x^j )$, be the ``spatial projection". The elements (points) of the space manifold $\mathrm{M}_\mathrm{F}$ are the {\it world lines}, each of which is the set of events that have a given spatial projection ${\bf x}$ in some chart $\chi: \mathrm{U} \rightarrow \mathbb{R}^4,\ X\mapsto {\bf X}$, belonging to the class $\mathrm{F}$. I.e., a world line $l$ is an element of $\mathrm{M}_\mathrm{F}$, iff there is a chart $\chi \in \mathrm{F}$ and a triplet ${\bf x} \in P_S(\chi (\mathrm{U}))$, such that $l$ is the set of all events $X$ in the domain U, whose spatial coordinates are ${\bf x}$:
\be\label{l-in-M-by-P_S}
l := \{\,X\in \mathrm{U};\ P_S(\chi (X))={\bf x}\,\}.
\ee
It results easily from (\ref{purely-spatial-change}) that (\ref{l-in-M-by-P_S}) holds true then in any chart $\chi' \in \mathrm{F}$, of course with the transformed spatial projection triplet ${\bf x}'= {\bf f}({\bf x}):= (f ^j({\bf x}))$ \cite{A44}. For any chart $\chi \in \mathrm{F}$, one defines the ``associated chart" as the mapping which associates, with a world line $l\in \mathrm{M}_\mathrm{F}$, the constant triplet of the spatial coordinates of the events $X\in l$:
\be\label{def-chi-tilde}
\widetilde{\chi }: \mathrm{M}\rightarrow \mathbb{R}^3,\quad l\mapsto {\bf x} \mathrm{\ such\ that\ }\forall X \in l,\  P_S(\chi (X))={\bf x}.\\
\ee 
The set \Couleur{$\mathrm{M}_\mathrm{F}$} is endowed with a natural structure of three-dimensional {\it differentiable manifold,} of which the basic atlas is made of the associated charts $\widetilde{\chi }$, where $\chi $ is any chart belonging to the reference frame $\mathrm{F}$ \cite{A44}. The ``spatial" objects defined above: the scalar \Couleur{$w$}, the vectors \Couleur{$\Mat{\Phi }$}, \Couleur{$\Mat{P}$}, and the tensor \Couleur{$\Mat{\Sigma }$}, are simply and rigorously tensor fields on the manifold \Couleur{$\mathrm{M}_\mathrm{F}$}. (Of course they have in general, in addition, a dependence on the time coordinate $x^0$, thus they are, strictly speaking, one-parameter families of tensor fields on \Couleur{$\mathrm{M}_\mathrm{F}$}.) Fixing a reference frame in this sense can be done, for instance, by choosing one local coordinate system (chart $\chi $) on the spacetime, with its domain of definition $\mathrm{U}$: the corresponding reference frame \Couleur{$\mathrm{F}$} is then the equivalence class of this chart. As soon as one has fixed a reference frame, then Eqs. (\ref{Local Conserv E-2}) and (\ref{Local Conserv P}), as well as Eqs. (\ref{Conserv E}) and (\ref{Conserv P}), are coordinate-free equations on the space manifold \Couleur{$\mathrm{M}_\mathrm{F}$}. In particular, the bounded spatial domain \Couleur{$\Omega $} is an open subset of the manifold \Couleur{$\mathrm{M}_\mathrm{F}$}, having a regular boundary \Couleur{$\partial \Omega $}, so that the divergence theorem applies. (See Appendix \ref{Domains} for a precise definition of the needed regularity.) \\

On the other hand, if one makes a general coordinate change, for which the change in the spatial coordinates depends on the time coordinate (already if one makes a Lorentz transformation transforming the Cartesian system into another one, but with a non-zero ``boost"), then (\ref{Def w Phi}) defines completely different quantities \Couleur{$w'$} and \Couleur{$\Mat{\Phi}'$}, as compared with the initial ones. The same is true for \Couleur{$\Mat{P}$} and \Couleur{$\Mat{\Sigma }$} as defined by (\ref{Def P Sigma}). This means that {\it there is one definition of the energy and momentum (and their fluxes) per reference frame.} It is not specific to special relativity. Indeed the energy depends on the reference frame. This is true in non-relativistic physics (e.g. \cite{A15}) --- as may be checked here on the fact that \Couleur{$w_\mathrm{m}$} and \Couleur{$\Mat{\Phi} _\mathrm{m}$} defined in Eqs. (\ref{w_m NG})--(\ref{Phi_m NG}) involve the velocity ${\bf v}$ that depends on the inertial frame, whereas \Couleur{$\rho , \Pi, U\ \mathrm{and}\ \Mat{\sigma }$} are Galilean invariants. It is also true in relativistic physics, and also in a general spacetime, and be it for the classical or the quantum-mechanical energy \cite{A51}.

\section{Definition of the energy-momentum tensor from a Lagrangian}\label{T-tensor}

\subsection{Lagrangian and stationary action principle}

\noi We assume that the equations of motion for some ``matter fields" \Couleur{$\phi ^A \ (A=1,...,n)$} derive from a Lagrangian \Couleur{$L$} through the {\it principle of stationary action} in a general spacetime:

\be\label{delta_S=0}
\operatorname{For\ any\ variation\ field\ }\Couleur{\delta \phi ^A=\delta \phi ^A( X)}\ \operatorname{with}\ \Couleur{\delta \phi ^A_{\ \mid \partial \mathrm{U}}=0},\ \operatorname{we\ have\ }\Couleur{\delta S=0}. 
\ee
Here, \Couleur{$\partial \mathrm{U}$} is the boundary, assumed smooth, of some bounded open set \Couleur{$\mathrm{U}$} in the spacetime, and \Couleur{$S$} is the action: in some chart \Couleur{$\chi $} whose domain of definition \Couleur{$\mathrm{W}$} contains \Couleur{$\mathrm{U}$}, it writes
\be\label{S general}
\Couleur{S = S_\mathrm{U}:= 
\int_{\chi (\mathrm{U})} L(\Mat{\phi }^A({\bf X}),\Mat{\phi }^A_{,\mu }({\bf X}),{\bf X}) \sqrt{-g({\bf X})}\, \dd ^4 {\bf X} },
\ee
where \Couleur{$ \Mat{\phi}^A:\ {\bf X} \mapsto \Mat{\phi}^A({\bf X}),\ \chi (\mathrm{W}) \rightarrow \mathbb{R}^{n_A}$}, is the local expression of the field \Couleur{$\phi ^A$} in the chart \Couleur{$\chi $}, and \Couleur{$g:=\mathrm{det}\,(g_{\mu \nu })$}, the \Couleur{$g_{\mu \nu }$} 's being the components of the metric tensor in the chart \Couleur{$\chi $} ; note that  \Couleur{$\chi (\mathrm{W})$} is an open subset of  \Couleur{$\mathbb{R}^4$}. Thus, the field \Couleur{$\phi ^A$} has \Couleur{$n_A$} real components (or \Couleur{$n_A$} complex components for a complex field, with \Couleur{$\mathbb{R}^{n_A}$} replaced by \Couleur{$\mathbb{C}^{n_A}$}). At this stage we do not need to know the exact geometric nature of the fields: whether they are scalars, vectors, more general tensors, or otherwise. We just assume that, on changing the chart: \Couleur{$\chi \hookrightarrow \chi '$}, the local expression of each of them has some definite transformation law, say \Couleur{$ \Mat{\phi}^A({\bf X}) \hookrightarrow \Mat{\phi}'^A({\bf X}')$},
\footnote{\ 
If \Couleur{$ \phi^A$} is a section of a vector bundle with base \Couleur{$\mathrm{V}$}, say \Couleur{$\mathrm{E}$}, to write its local expression needs that not only a chart on \Couleur{$\mathrm{V}$} but also a frame field \Couleur{$(e_a)$} on \Couleur{$\mathrm{E}$} is given. However, in the case of a tensor field, a relevant frame field is determined uniquely by the data of the chart \Couleur{$\chi $} with the associated natural basis \Couleur{$(\partial _\mu ) $} and the dual basis \Couleur{$(\dd x^\mu ) $}. In a very general case, \Couleur{$\mathrm{E}$} has the form \Couleur{$\mathrm{E}=\mathrm{T}\otimes \mathrm{N}$} with \Couleur{$\mathrm{T}$} an usual tensor bundle on \Couleur{$\mathrm{V}$} and \Couleur{$\mathrm{N}$} a vector bundle of a different kind. Then it is natural to take a frame field of the form \Couleur{$(T_b\otimes N_c)$}, with \Couleur{$(T_b)$} a frame field on \Couleur{$\mathrm{T}$}, determined by the chart \Couleur{$\chi $}, and with \Couleur{$(N_c)$} a frame field on \Couleur{$\mathrm{N}$}, which is left unchanged when changing the chart. Thus the transformation law on changing the chart is determined.
}
and that the Lagrangian is then invariant under the coordinate change: 
\be\label{L invariant}
\Couleur{L(\Mat{\phi }'^A({\bf X}'),\Mat{\phi }'^A_{,\mu }({\bf X}'),{\bf X}') = L(\Mat{\phi }^A({\bf X}),\Mat{\phi }^A_{,\mu }({\bf X}),{\bf X}), \quad {\bf X}'=\chi '(\chi ^{-1}({\bf X})):=F({\bf X})}.
\ee
That invariance has to be true at least when the chart belongs to some well-defined class and implies that the same invariance is valid for the action (\ref{S general}). In this section, we shall consider the usual case that {\it all} charts (in the atlas of the spacetime manifold) are allowed, i.e., we shall discuss generally-covariant theories. However, it also makes sense to consider instead the class associated with a particular (``privileged") reference frame. Thus, the Lagrangian is a smooth real function \Couleur{$L=L({\bf q} ^A, {\bf q} ^A_\mu ,{\bf X} )$}, where \Couleur{${\bf X}\in \mathbb{R}^4$} is the coordinate vector specifying the spacetime position, \Couleur{${\bf q}^A \in \mathbb{R}^{n_A}$}, and also \Couleur{$ {\bf q}^A_\mu \in \mathbb{R}^{n_A}$} for \Couleur{$\mu=0,...,3$}. These five vectors of \Couleur{$\mathbb{R}^{n_A}$}  specify the values that may be taken at \Couleur{${\bf X}$} by the local expression of the field \Couleur{$\phi ^A$} and its partial derivatives. This means that, in the expression (\ref{S general}) of the action, one makes the substitution
\be\label{Assign q^A q^A_mu}
\Couleur{{\bf q}^A=\Mat{\phi} ^A({\bf X}),\quad {\bf q}^A_\mu  =\Mat{\phi} ^A_{,\mu  }({\bf X}) \quad (A=1,...,n;\ \mu =0,...,3)}.
\ee
Note also that \Couleur{$\dd V_4 :=\sqrt{-g}\dd ^4 {\bf X}$}, with \Couleur{$g:=\mathrm{det}\,(g_{\mu \nu })$}, is the invariant four-volume element on the spacetime [thus \Couleur{$g<0$} for a Lorentzian metric on the four-dimensional spacetime
].\\
 
The stationarity (\ref{delta_S=0}) is equivalent to the Euler-Lagrange equations \{see e.g. \cite{L&L,Olver2012}; \Couleur{$\delta S$} is defined from a Gateaux derivative, as with Eq. (\ref{Def delta S}) below\}. In a general spacetime, the latter equations write \cite{Leclerc2006b}:
\be\label{E-L general}
\Couleur{\partial _\mu \left(\frac{\partial \mathcal{L}}{\partial {\bf q} ^A_{\mu }} \right) = \frac{\partial \mathcal{L}}{\partial {\bf q} ^A} \quad (A=1,...,n),\qquad \mathcal{L}:= L \sqrt{-g}      }
\ee
in \Couleur{$\mathrm{U}$}, with the implicit assignment (\ref{Assign q^A q^A_mu}). The domain of definition \Couleur{$\mathrm{W}$} of the coordinate system now has to contain not only \Couleur{$\mathrm{U}$} but also the boundary \Couleur{$\partial \mathrm{U}$}, because the derivation of (\ref{E-L general}) needs to use the divergence theorem. 

\subsection{The ``canonical" (or ``Noether") T-tensor}

We shall give only a very brief account (see e.g. \cite{L&L, Belinfante1939,GotayMarsden1992,ForgerRoemer2004,Leclerc2006b}). This object has the following expression in a given chart:
\be\label{tau tensor}
\Couleur{\tau _\mu ^{\ \,\nu }({\bf X})=\Mat{\phi} ^A_{,\mu }({\bf X})\left(\frac{\partial L}{\partial {\bf q} ^A_{\nu }}\right)_{{\bf q}^B=\Mat{\phi} ^B({\bf X}),\ {\bf q}^B_\rho =\Mat{\phi} ^B_{,\rho }({\bf X})}-\delta _\mu ^\nu L(\Mat{\phi }^A({\bf X}),\Mat{\phi }^A_{,\rho  }({\bf X}),{\bf X})}.
\ee
When \Couleur{$-g=1$} and \Couleur{$\mathcal{L}=L=L({\bf q} ^A, {\bf q} ^A_{\mu })$} does not depend explicitly on the spacetime position, this object occurs naturally from the derivation of the Euler-Lagrange equations (\ref{E-L general}), which imply that it verifies the desired local conservation equation \Couleur{$\ \tau  _{\mu \ \,,\nu }^{\ \nu }=0$} \cite{L&L}. However, such an independence happens in practice only in a flat spacetime. Moreover, in fact, this object is not necessarily a tensor --- even in a flat spacetime, cf. the case of the electromagnetic field \cite{Leclerc2006b}: \Couleur{$\Mat{F}$} being the field tensor and \Couleur{$\Mat{A}$} the 4-potential, we have \cite{L&L,GotayMarsden1992} 
\be\label{Canonical_em}
\Couleur{4\pi \tau _\mu ^{\ \,\nu }=-A_{\rho ,\mu }F^{\nu \rho }+\frac{1}{4}(F_{\rho \sigma }F^{\rho \sigma })\delta _\mu ^\nu}. 
\ee
(Henceforth, indices are raised or lowered with the spacetime metric.) On the r.h.s., everything, {\it but} \Couleur{$A_{\rho ,\mu }$}, is tensorial, hence \Couleur{$\tau _\mu ^{\ \,\nu }$} is not a tensor; i.e., \Couleur{$\tau _\mu ^{\ \,\nu }$} does not transform as a $(1\ 1)$ tensor for general coordinate changes. Of course this does not mean that there is no energy-momentum tensor for the electromagnetic field. (The Hilbert tensor indeed does the job, see e.g. \cite{L&L}.) But it proves that the ``canonical tensor" is not necessarily a tensor. This is not often noted, e.g. it is not in Refs. \cite{L&L,GotayMarsden1992}, probably because (\ref{Canonical_em}) does behave as a $(1\ 1)$ tensor for linear coordinate changes, as are the Lorentz transformations to which one often restricts oneself in special relativity. In a general spacetime, \Couleur{$\Mat{\tau }$} is a tensor for a scalar field \cite{Leclerc2006b} --- and also for the Dirac field \cite{A48}.

\subsection{Hilbert's variational definition of the T-tensor}\label{Hilbert Tensor}

While following the line of the classic derivation by Landau \& Lifshitz \cite{L&L}, we will include many mathematical details which appear necessary in that derivation and that, for the most part, we did not find in the literature that we consulted. By this, we do not mean the geometric formulation of the physical fields as sections of appropriate fibre bundles, which has been implemented in Ref. \cite{GotayMarsden1992}, among others, and rather extensively in Ref. \cite{ForgerRoemer2004} --- and which we will not need. [See the remarks following the definition (\ref{S general}) of the action.] Instead, we mean the precise definition of ``the variation of the action under an infinitesimal diffeomorphism", the regularity of the boundary and the exact boundary conditions, and a clear derivation of the main formulas.\\

One considers a given chart \Couleur{$\chi :X\mapsto {\bf X}=(x^\mu )$} and one imposes a small change to it: \Couleur{$x^\mu \hookrightarrow x^\mu +\delta x^\mu $}. As we shall see, the domain of definition \Couleur{$\mathrm{W}$} of \Couleur{$\chi$} must include the closure \Couleur{$\overline{\mathrm{U}}$} of the bounded open set \Couleur{$\mathrm{U}$} in which one computes the action (\ref{S general}), and we must assume that \Couleur{$\delta x^\mu=0 $} at the events \Couleur{$X \in\mathrm{W}$} which do not belong to \Couleur{$\mathrm{U}$}. (Alternatively, one may regard the mapping defined in coordinates by \Couleur{$x^\mu \mapsto x^\mu +\delta x^\mu $} as a diffeomorphism of the spacetime manifold  \Couleur{$\mathrm{V}$}, which coincides with the identity map for \Couleur{$X \notin \mathrm{U}$}.) Thus, \Couleur{$\delta x^\mu=\epsilon \xi ^\mu $} with \Couleur{$\xi$} any smooth vector field that vanishes if \Couleur{$X \notin \mathrm{U}$}, and \Couleur{$\epsilon \ll 1$}. That is, we change the chart \Couleur{$\chi $} for a new chart \Couleur{$\chi_\epsilon  $} given by
\be\label{Def chi_epsilon}
\Couleur{\chi _\epsilon (X):={\bf X}+\epsilon \Mat{\xi}({\bf X})} \quad \Couleur{X\in \mathrm{W}},\quad \Couleur{{\bf X}:=\chi (X)}.
\ee
(The vector field \Couleur{${\bf X} \mapsto \Mat{\xi}({\bf X})=(\xi ^\mu ({\bf X}))$} is the local expression of \Couleur{$\xi$} in the chart \Couleur{$\chi$}.) After such a coordinate change, the local expressions of the fields change, each according to its specific transformation behaviour, and the domain \Couleur{$\chi (\mathrm{U})\subset \mathbb{R}^4$} also changes, so both the integrand and the integration domain change in the action (\ref{S general}), which thus takes {\it a priori} a different value. One seeks to calculate the first-order term, as \Couleur{$\epsilon \rightarrow 0$}, in the variation of the action, \Couleur{$S_\mathrm{U}$} being considered as a functional of the field \Couleur{${\bf X} \mapsto \delta {\bf X} =\epsilon \Mat{\xi} ({\bf X}) $}. This amounts to calculate the Gateaux derivative of \Couleur{$S_\mathrm{U}$} in the direction of the vector field \Couleur{$\Mat{\xi}$}:
\be\label{Def delta S}
\Couleur{\frac{(\delta S_\mathrm{U})(\epsilon )}{\epsilon }:= \left(\frac{\dd }{\dd \epsilon }S_\mathrm{U}(\epsilon \Mat{\xi})\right)_{\epsilon =0}}.
\ee
The bounded open set \Couleur{$\mathrm{U}$} is assumed to be an {\it open domain} in the sense of \hyperref[Proposition 2]{Proposition 2} in Appendix \ref{Domains}, in particular it has a smooth boundary \Couleur{$\partial \mathrm{U}$}. Because \Couleur{$\mathrm{U}$} is an open domain, the assumption ``\Couleur{$\xi(X)=0 $} if \Couleur{$X \notin \mathrm{U}$}" means exactly that the support of \Couleur{$\xi $} is included in the closure \Couleur{$ \overline{\mathrm{U}} = \mathrm{U}\cup \partial \mathrm{U}$} (\hyperref[Corollary]{Corollary} to \hyperref[Proposition 2]{Proposition 2}). 
\footnote{\ \label{K vs U}
Considering ``every compact subset \Couleur{$\mathrm{K}$} of \Couleur{$\mathrm{V}$}" \cite{ForgerRoemer2004} is too general, because one needs to use the divergence theorem on \Couleur{$\mathrm{K}$} to eliminate a surface term, as is done below to get Eqs. (\ref{d S_epsilon/d epsilon-2}) and (\ref{delta_S_coordinate-change}), and as is done for the equation after Eq. (222) in Ref. \cite{ForgerRoemer2004}. One should assume that the compact set is a ``closed domain". A closed domain \Couleur{$\mathrm{D}$} is the closure of its interior \Couleur{$\interior{\mathrm{D}}$} (\hyperref[Proposition 2]{Proposition 2}), hence it is equivalent to start from a bounded open domain \Couleur{$\mathrm{U}$} and to define \Couleur{$\mathrm{K} = \overline{\mathrm{U}}$}, which is a compact domain, or to start from \Couleur{$\mathrm{K}$} and to define \Couleur{$\mathrm{U}=\interior{\mathrm{K}}$}. Then, of course, the smooth boundary \Couleur{$\partial \mathrm{U}$} has measure zero with respect to the invariant four-dimensional measure \Couleur{$\sqrt{-g}\,\dd ^4 {\bf X}$}. Hence, the action integral (\ref{S general}), as well as its variation (\ref{delta_S_coordinate-change}) or (\ref{delta_S_coordinate-change-T}), are unaltered if instead of \Couleur{$\mathrm{U}$}  we consider its closure \Couleur{$\mathrm{K}$} as the integration domain, as considered in Ref. \cite{ForgerRoemer2004}. Thus, when the compact \Couleur{$\mathrm{K}$} considered in Ref. \cite{ForgerRoemer2004} is a closed domain, as it should be imposed, the definition of the action \Couleur{$S_\mathrm{K}$} and the condition imposed on the vector field \Couleur{$X$} in Ref. \cite{ForgerRoemer2004} are equivalent to those considered here for  \Couleur{$S_\mathrm{U}$} and \Couleur{$\xi$}.
}
Since \Couleur{$\mathrm{Supp} (\partial _\nu \xi ^\mu )\subset \mathrm{Supp} (\xi ^\mu )$}, this implies that all derivatives of \Couleur{$\xi $} also vanish if \Couleur{$X \notin \mathrm{U}$}. It follows that the corresponding change in \Couleur{$g^{\mu \nu }$}, as determined to the first order in \Couleur{$\epsilon $} \{Eq. (94.2) in Ref. \cite{L&L}\}:
\be\label{delta g}
\Couleur{\delta g^{\mu \nu }=\epsilon \left(\xi ^{\mu ;\nu }+\xi ^{\nu ;\mu } \right ) = \delta x ^{\mu ;\nu }+\delta x ^{\nu ;\mu }}, 
\ee
also vanishes if \Couleur{$X \notin \mathrm{U}$}. [Here, the \Couleur{$g^{\mu \nu }$} 's are the components of the inverse of the metric's component matrix \Couleur{$(g_{\mu \nu })$}.] Thus, in particular, \Couleur{$\delta g^{\mu \nu }=0$} on \Couleur{$\partial \mathrm{U}$}. 
\footnote{\ 
In Ref. \cite{L&L}, it is ``set \Couleur{$\delta g^{\mu \nu }=0$} at the integration limits" and, later on, it is stated that ``the  \Couleur{$\xi ^\mu $} vanish at the limits of the integration", without any explanation nor any link between the two statements. Note that assuming merely that \Couleur{$\xi ^\mu =0$} on the boundary \Couleur{$\partial \mathrm{U}$} does {\it not} imply that \Couleur{$\delta g^{\mu \nu }=0$} on \Couleur{$\partial \mathrm{U}$}. 
}

One assumes moreover that the matter Lagrangian \Couleur{$L$} depends on the matter fields and the metric and their first-order derivatives, but not on their higher-order derivatives. The same applies then to the ``Lagrangian density" \Couleur{$\mathcal{L}:=L\sqrt{-g}$}, i.e., the latter is a smooth real function \Couleur{$\mathcal{L}=\mathcal{L}({\bf q}^A, {\bf q}^A_{\mu },\underline{g}^{\mu \nu },\underline{g}^{\mu \nu }_{\rho })$}. To calculate the action (\ref{S general}), in the Lagrangian \Couleur{$L$} the values (\ref{Assign q^A q^A_mu}) are assigned to \Couleur{${\bf q}^A$} and \Couleur{${\bf q}^A_{\mu }$}. In addition, now one assigns the values \Couleur{$g^{\mu \nu }({\bf X})$} and \Couleur{$g^{\mu \nu }_{\ \, ,\rho }({\bf X})$} to \Couleur{$\underline{g}^{\mu \nu }$} and \Couleur{$\underline{g}^{\mu \nu }_{\rho }$} respectively. At this stage, it is usually admitted that the change in the action (\ref{S general}), \Couleur{$\delta S_\mathrm{U}$}, is given by
\be\label{delta_S_coordinate-change-0}
\Couleur{\delta S_\mathrm{U}=\int_{\chi(\mathrm{U})}  \left [\frac{\partial \mathcal{L}}{\partial {\bf q}^A} \delta \Mat{\phi }^A +\frac{\partial \mathcal{L}}{\partial ({\bf q}^A_{\mu })} \delta (\Mat{\phi }^A_{,\mu }) + \frac{\partial \mathcal{L}}{\partial \underline{g}^{\mu \nu }}\delta g^{\mu \nu }+ \frac{\partial \mathcal{L}}{\partial \left(\underline{g}^{\mu \nu }_{\rho }\right)} \delta (g^{\mu \nu }_{\ \, ,\rho })\right ] \, \dd ^4 {\bf X}},
\ee
without a justification nor a precise definition of \Couleur{$\delta S_\mathrm{U}$}, \Couleur{$\delta \Mat{\phi }^A$}, etc. \{Actually, the step (\ref{delta_S_coordinate-change-0}) and some later intermediate steps are even skipped by Landau \& Lifshitz \cite{L&L}.\} However, recall that here not only the integrand but also the integration domain change in the action (\ref{S general}), so the first-order variation of the integral \Couleur{$S_\mathrm{U}$} involves {\it a priori} more than just the first-order approximation of the integrand. Let us be more precise. Denote provisionally the list of all fields (matter fields and metric) by \Couleur{$\phi ^J \ (J=1,...,n+1)$}. Applying the definition (\ref{S general}) with the new chart (\ref{Def chi_epsilon}), we write 
\be\label{S_epsilon} 
\Couleur{S_\epsilon :=S_\mathrm{U}(\epsilon \Mat{\xi}) = \int_{F_\epsilon (\Omega  )} \widetilde{\mathcal{L}}({\bf X},\epsilon ) \dd^4 {\bf X}},
\ee
with \Couleur{$\Omega:=\chi (\mathrm{U})$},
\be
\Couleur{F_\epsilon :=\chi _\epsilon \circ \chi ^{-1}},
\ee
\be
\Couleur{\widetilde{\mathcal{L}}({\bf X},\epsilon ) := \mathcal{L}(\Mat{\phi} ^J_\epsilon({\bf X}),\partial _\mu \Mat{\phi} ^J_\epsilon({\bf X}))},
\ee
where \Couleur{$\Mat{\phi} ^J_\epsilon: \chi _\epsilon (\mathrm{W}) \rightarrow \mathbb{R}^{n_J}$} is the local expression of the field \Couleur{$\phi ^J$} in the chart \Couleur{$\chi _\epsilon $}. The integral (\ref{S_epsilon}) has a form that is well known, in particular, in continuum mechanics. The expression of its derivative is also well known:
\be\label{d S_epsilon/d epsilon}
\Couleur{\frac{\dd S_\epsilon }{\dd \epsilon }= \int_{F_\epsilon (\Omega  )} \left[\frac{\partial \widetilde{\mathcal{L}}}{\partial \epsilon } +\mathrm{div}(\widetilde{\mathcal{L}}  \Mat{\xi }_\epsilon )\right] \dd^4 {\bf X}},
\ee
where \Couleur{$\Mat{\xi }_\epsilon({\bf X}) := \frac{\partial F_\epsilon }{\partial \epsilon }\left({\bf X}_0\right)$}, with  \Couleur{${\bf X}_0:= F_\epsilon^{-1} ({\bf X})$}, is the ``velocity field" at ``time" \Couleur{$\epsilon $}. In particular, we get from (\ref{Def chi_epsilon}) and these definitions that \Couleur{$\Mat{\xi }_{\epsilon=0} =\Mat{\xi }$}. Therefore, applying the divergence theorem in (\ref{d S_epsilon/d epsilon}) and since \Couleur{$\xi=0$} on \Couleur{$\partial \mathrm{U}$} (hence \Couleur{$\Mat{\xi }={\bf 0}$} on \Couleur{$\partial \Omega $}, because \Couleur{$\partial \Omega = \partial\, (\chi (\mathrm{U}))=\chi (\partial \mathrm{U}) $}), we get
\bea\label{d S_epsilon/d epsilon-2}
\Couleur{\left(\frac{\dd S_\epsilon }{\dd \epsilon }\right) _{\epsilon =0}} & \Couleur{=} & \Couleur{\int_{\Omega  } \left( \frac{\partial \widetilde{\mathcal{L}}}{\partial \epsilon } \right) _{\epsilon =0} \dd^4 {\bf X} } \\
& = & \Couleur{\int_{\Omega  } \left[\frac{\partial \mathcal{L}}{\partial {\bf q}^J} \left(\frac{\partial (\Mat{\phi} ^J_\epsilon({\bf X}))}{\partial \epsilon }\right)_{\epsilon =0} + \frac{\partial \mathcal{L}}{\partial {\bf q}^J_\mu } \left(\frac{\partial (\partial _\mu \Mat{\phi} ^J_\epsilon({\bf X}))}{\partial \epsilon }\right)_{\epsilon =0} \right] \dd^4 {\bf X}.} \nonumber 
\eea
Hence, defining simply the variations of the fields to the first order in \Couleur{$\epsilon$}:
\be\label{Def dphi dphi_mu}
\Couleur{\delta \Mat{\phi} ^J({\bf X}):=\epsilon \left(\frac{\partial (\Mat{\phi} ^J_\epsilon({\bf X}))}{\partial \epsilon }\right)_{\epsilon =0}, \quad \delta \left(\Mat{\phi} ^J_{,\mu}\right) ({\bf X}):=\epsilon \left(\frac{\partial (\partial _\mu \Mat{\phi} ^J_\epsilon({\bf X}))}{\partial \epsilon }\right)_{\epsilon =0}},
\ee
and remembering that the \Couleur{$\Mat{\phi} ^J$} 's are the matter fields \Couleur{$\Mat{\phi} ^A$} for \Couleur{$J=A=1,...,n$}, while \Couleur{$\Mat{\phi} ^{n+1}$} is the metric field \Couleur{$g^{\mu \nu }$}, we see that Eq. (\ref{d S_epsilon/d epsilon-2}), together with the definition (\ref{Def delta S}), proves the ``obvious" Eq. (\ref{delta_S_coordinate-change-0}). We see then from Eq. (\ref{d S_epsilon/d epsilon}) that Eq. (\ref{delta_S_coordinate-change-0}) is in general {\it false} if the boundary condition \Couleur{$\Mat{\xi }={\bf 0}$} on \Couleur{$\partial \Omega $} is not valid. \\

If we write
\be
\Couleur{\left(\frac{\partial (\Mat{\phi} ^J_\epsilon({\bf X}))}{\partial \epsilon }\right)_{\epsilon =0} =\lim_{\epsilon \rightarrow 0} \frac{\Mat{\phi} ^J_\epsilon({\bf X})-\Mat{\phi} ^J_0({\bf X})}{\epsilon }},
\ee
and if we remember that \Couleur{$\Mat{\phi} ^J_\epsilon$} is the local expression of the field \Couleur{$\phi ^J$} in the chart (\ref{Def chi_epsilon}), that follows the flow of the vector field \Couleur{$\xi $} at small values of \Couleur{$\epsilon $}, we recognize (at least in the case that \Couleur{$\phi ^J$} is a tensor field) the definition of the Lie derivative \cite{Doubrovine1982} --- or rather, of its opposite. That is,
\be\label{d phi Lie}
\Couleur{\delta \Mat{\phi} ^J({\bf X}) =-\epsilon L_\xi \Mat{\phi} ^J({\bf X})}.
\ee
We note also that
\be\label{d_mu dphi/deps =d/deps d_mu phi}
\Couleur{\frac{\partial (\partial _\mu \Mat{\phi} ^J_\epsilon({\bf X}))}{\partial \epsilon } = \partial _\mu \left(\frac{\partial (\Mat{\phi} ^J_\epsilon({\bf X}))}{\partial \epsilon }\right)},
\ee
hence
\be\label{dphi_mu = d_mu dphi}
\Couleur{ \delta \left(\Mat{\phi} ^J_{,\mu}\right) ({\bf X})=\partial _\mu (\delta \Mat{\phi} ^J({\bf X}))}.
\ee
In practical terms, \Couleur{$\delta \Mat{\phi} ^J({\bf X})$} can be computed in two steps \cite{L&L}: first, one computes the difference \Couleur{$\Mat{\phi}^J_\epsilon ({\bf X}')-\Mat{\phi}^J({\bf X})$} between the local expressions in the initial and the modified chart at points that {\it correspond} together through the transition map, i.e. \Couleur{${\bf X}'=F_\epsilon ({\bf X})$}. Second, one uses a first-order Taylor expansion to find the difference \Couleur{$\Mat{\phi}^J_\epsilon ({\bf X})-\Mat{\phi}^J ({\bf X})$}, i.e., at one and the same point. For instance, this gives indeed the expression (\ref{delta g}) for the metric \cite{L&L}. As another example, consider a vector field \Couleur{$V$}, with components \Couleur{$V^\mu $} in the initial chart. We find without difficulty
\be\label{delta V^mu}
\Couleur{\delta V^\mu = \delta x^\mu _{,\nu }V^\nu -V^\mu _{,\nu }\delta x^\nu = \delta x^\mu _{;\nu }V^\nu -V^\mu _{;\nu }\delta x^\nu},
\ee
where the second equality occurs due to the symmetry of the Christoffel symbols (i.e., due to the fact that the Levi-Civita connection has no torsion). This does coincide with the Lie derivative definition (\ref{d phi Lie}).\\

In a last step, let us assume that the matter fields obey the Euler-Lagrange equations (\ref{E-L general}). Just like for the derivation of the latter equations from the stationarity of the action (\ref{S general}): because \Couleur{$\delta (\Mat{\phi }^A_{,\mu })=(\delta \Mat{\phi }^A)_{,\mu }$} in view of (\ref{dphi_mu = d_mu dphi}), one may transform the second term in (\ref{delta_S_coordinate-change-0}) and use the divergence theorem to make a surface term appear in it; and that surface term vanishes because \Couleur{$\delta \Mat{\phi }^A=0$} on \Couleur{$\partial \mathrm{U}$} [as can be checked for a vector on (\ref{delta V^mu}) and for the contravariant metric tensor on (\ref{delta g})]. It then follows from (\ref{E-L general}) that the first two terms in (\ref{delta_S_coordinate-change-0}) cancel one another, thus
\footnote{\label{g_mu nu independent}\ 
All sixteen \Couleur{$\underline{g}^{\mu \nu}$} 's (\Couleur{$0 \leq \mu \leq 3, \ 0\leq \nu \leq 3$}) are considered as independent variables in \Couleur{$\mathcal{L}$} for the calculation of \Couleur{$\partial \mathcal{L}/\partial \underline{g}^{\mu \nu }$}, even though \Couleur{$g^{\mu \nu}=g^{\nu \mu}$}: see Note $\ddagger $ on p. 269 in Ref. \cite{L&L}. Thus, all sixty-four \Couleur{$\underline{g}^{\mu \nu }_{\rho }$} 's (\Couleur{$0 \leq \mu \leq 3, \ 0\leq \nu \leq 3, \ 0\leq \rho \leq 3$}) are also considered as independent variables for the calculation of \Couleur{$\partial \mathcal{L}/\partial \underline{g}^{\mu \nu }_\rho $}.
}
\be\label{delta_S_coordinate-change-1}
\Couleur{\delta S_\mathrm{U}=\int_{\chi(\mathrm{U})} \left [\frac{\partial \mathcal{L}}{\partial \underline{g}^{\mu \nu }}\delta g^{\mu \nu }+ \frac{\partial \mathcal{L}}{\partial \left(\underline{g}^{\mu \nu }_{\rho }\right)} \delta (g^{\mu \nu }_{\ \, ,\rho })\right ] \, \dd ^4 {\bf X}}, \qquad \Couleur{\mathcal{L}:= L \sqrt{-g} }. 
\ee
In the same way, because \Couleur{$\delta (g^{\mu \nu }_{\ \, ,\rho })=(\delta g^{\mu \nu })_{,\rho  }$}, one may transform the second term in (\ref{delta_S_coordinate-change-1}) and use the divergence theorem to make a surface term appear in it, and that surface term vanishes because \Couleur{$\delta g^{\mu \nu }=0$} on \Couleur{$\partial \mathrm{U}$}. One thus gets \cite{L&L}:
\be\label{delta_S_coordinate-change}
\Couleur{\delta S_\mathrm{U}=\int_{\chi(\mathrm{U})} \left [\frac{\partial \mathcal{L}}{\partial \underline{g}^{\mu \nu }}-\frac{\partial }{\partial x^\rho} \left ( \frac{\partial \mathcal{L}}{\partial \left(\underline{g}^{\mu \nu }_{\rho }\right)}\right )\right ] \delta g^{\mu \nu }\, \dd ^4 {\bf X}}. 
\ee
We have thereby proved the following:
\paragraph{Theorem 1.}\label{Theorem 1} {\it Assume that the bounded subset \Couleur{$\mathrm{U}$} of the spacetime is an open domain in the sense of \hyperref[Proposition 2]{Proposition 2}. In the domain \Couleur{$\mathrm{W}\supset \overline{\mathrm{U}}$} of some chart \Couleur{$\chi$}, define a one-parameter family of new charts by (\ref{Def chi_epsilon}), associated with a smooth vector field \Couleur{$\xi$} that is defined over \Couleur{$\mathrm{W}$} and vanishes for \Couleur{$X \notin \mathrm{U}$}. Assume that the smooth matter Lagrangian has the form \Couleur{$L=L({\bf q}^A, {\bf q}^A_{\mu },\underline{g}^{\mu \nu },\underline{g}^{\mu \nu }_{\rho })$}. Then:}\\
(i) {\it The first-order variation \Couleur{$\delta S_\mathrm{U}$} of the action, defined in Eq. (\ref{Def delta S}), is given by Eq. (\ref{delta_S_coordinate-change-0}). In this equation, the variations \Couleur{$\delta \Mat{\phi} ^J$} and \Couleur{$\delta \left(\Mat{\phi} ^J_{,\mu}\right)\ (J=1,...,n+1$}, with \Couleur{$\Mat{\phi} ^{n+1}=(g^{\mu \nu })$)} are defined by Eq. (\ref{Def dphi dphi_mu}) or equivalently by Eqs. (\ref{d phi Lie}) and (\ref
{dphi_mu = d_mu dphi}).} \\
(ii) {\it If the matter fields \Couleur{$\phi ^A \ (A=1,...,n)$} obey the Euler-Lagrange equations (\ref{E-L general}), then \Couleur{$\delta S_\mathrm{U}$} is given by Eq. (\ref{delta_S_coordinate-change}) right above. }\\

\noi Equation (\ref{delta_S_coordinate-change}) leads one to define an object  \Couleur{$\Mat{T}$} (usually called ``Hilbert energy-momentum tensor", though not in Ref. \cite{L&L}) by its components \cite{L&L}:
\be\label{Hilbert tensor}
\Couleur{\frac{1}{2}\sqrt{-g}\,T_{\mu \nu }:=  \frac{\partial \mathcal{L}}{\partial \underline{g}^{\mu \nu }}-\frac{\partial }{\partial x^\rho} \left ( \frac{\partial \mathcal{L}}{\partial \left(\underline{g}^{\mu \nu }_{\rho }\right)}\right), \qquad \Couleur{\mathcal{L}:= L \sqrt{-g} }.   }
\ee
The symmetry of this object: \Couleur{$T_{\nu \mu }=T_{\mu \nu }$}, follows from the symmetry of the metric {\it and} the invariance of \Couleur{$L$} under general coordinate changes. 
\footnote{\
Reminding Note (\ref{g_mu nu independent}), the symmetry of the metric does not by itself imply the symmetry of the components (\ref{Hilbert tensor}): check e.g. \Couleur{$L=(g^{12})^2$}, which of course is not invariant. 
}
As shown by \hyperref[Theorem 1]{Theorem 1}: For the object \Couleur{$\Mat{T}$} whose components are defined by Eq. (\ref{Hilbert tensor}), we have ``on shell" for any regular bounded open set \Couleur{$\mathrm{U}$} and for any coordinate change \Couleur{$\delta x^\mu=\epsilon \xi ^\mu $} such that \Couleur{$\xi^{\mu}(X)$} vanishes for \Couleur{$X \notin \mathrm{U}$}:
\be\label{delta_S_coordinate-change-T}
\Couleur{\delta S_\mathrm{U}=\frac{1}{2}\int_{\chi(\mathrm{U})} T_{\mu \nu }\,\delta g^{\mu \nu }\, \sqrt{-g}\,\dd ^4 {\bf X}}. 
\ee
If the Lagrangian function \Couleur{$L$} is {\it invariant under general coordinate changes,} then the action \Couleur{$S_\mathrm{U}$} in Eq. (\ref{S general}) is invariant too, hence the change \Couleur{$\delta S_\mathrm{U}$} given by Eq. (\ref{delta_S_coordinate-change}) or (\ref{delta_S_coordinate-change-T}) is zero for any possible coordinate change. Assume, moreover, that the object \Couleur{$\Mat{T}$} given by (\ref{Hilbert tensor}) turns out to be indeed a tensor. (This tensorial character does not seem to be proved in the literature, but see Subsect. \ref{Unique Hilbert tensor} below.) Then, using the expression (\ref{delta g}) of \Couleur{$\delta g^{\mu \nu }$} in terms of the vector field \Couleur{$\xi$}, and since by assumption the latter vanishes on \Couleur{$\partial \mathrm{U}$}, one gets from (\ref{delta_S_coordinate-change-T}) \cite{L&L}:
\be\label{Semicolon_Cons}
\Couleur{T _{\mu \ \ ;\nu }^{\ \,\nu} =0}.
\ee


\subsection{Is Eq. (\ref{Semicolon_Cons}) a true conservation equation?}
 
In contrast with (\ref{True_Cons}) [\Couleur{$T _{\mu \ \ ,\nu }^{\ \,\nu} =0$}, with partial derivatives], Eq. (\ref{Semicolon_Cons}) [with covariant derivatives] ``does not generally express any conservation law whatever" \cite{L&L}. Fock \cite{Fock1959} used similar words: he noted that the four scalar equations contained in (\ref{Semicolon_Cons}) ``do not by themselves lead to conservation laws". To explain it quickly, the presence of covariant derivatives gives to  Eq. (\ref{Semicolon_Cons}) the form of (\ref{True_Cons}) {\it plus source terms,} which are the terms linear in the T-tensor itself (that involve the connection coefficients). Nevertheless, Eq. (\ref{Semicolon_Cons}) can be rewritten in the form of (\ref{True_Cons}) after introducing some ``pseudo-tensor of the gravitational field" \Couleur{$\Mat{t}$}. But the definition of \Couleur{$\Mat{t}$} is not unique. And \Couleur{$\Mat{t}$} behaves as a tensor only for linear coordinate transformations. As a result, it is generally agreed that Eq. (\ref{Semicolon_Cons}) can lead only (under special assumptions, e.g. an asymptotically flat spacetime) to {\it global} conservation laws, see e.g. \cite{L&L,Stephani}. However, in order to be able to investigate the energy balance in any spatial domain, one would need to know uniquely the relevant energy density and its flux. And one would need that they obey a true and local conservation equation. (This is indeed the case in most fields of physics, e.g. in mechanics, thermodynamics, electrodynamics, chemistry, ..., as well as in Newtonian gravitation --- as shown in Sect. \ref{NG} --- and also in several alternative relativistic theories of gravitation in a flat spacetime, e.g. \cite{Rosen1963,LogunovMestvirishvili1989,Petry2014}, including a preferred-frame scalar theory with a mechanism for gravity \cite{A35}.)\\

What is thus lacking in theories based on Eq. (\ref{Semicolon_Cons}), which include general relativity and its numerous variants or extensions, is not merely an exact local concept of the {\it gravitational} energy. As we recalled, in special relativity the local conservation equation (\ref{True_Cons}) for the energy-momentum tensor  can be rewritten as two local conservation equations of the type (\ref{Local Conserv E}): a scalar one for the energy and a vector one for the three-momentum. We believe that the local concept of energy is indissolubly bound with the existence of a true local conservation equation of the type (\ref{Local Conserv E}), as it exists both in non-relativistic physics and in special relativity. Since, as we recalled, the rewriting of Eq. (\ref{Semicolon_Cons}) as an exact local conservation equation having the form (\ref{True_Cons}) is neither tensorial nor unique, we consider that Eq. (\ref{Semicolon_Cons}) does not provide an exact local concept for any form of {\it material energy,} either: assuming the definition of the Hilbert tensor  \Couleur{$\Mat{T}$} is unique (that is proved in Subsect. \ref{Unique Hilbert tensor}), one could define the material energy density as, say, \Couleur{$w:=T^{00}$}. But another one could prefer to choose \Couleur{$w'=T_{00}$}, and a third one, \Couleur{$w'':=T_0^0$}. In the absence of an exact local conservation equation of the type (\ref{True_Cons}), nobody can tell which choice is correct. Another point is of course that, for a given bounded spatial domain \Couleur{$\Omega $}, the time evolution of, say, the integral of \Couleur{$w:=T^{00}$} [the l.h.s. of Eq. (\ref{Conserv E})] is not equal to a flux through the boundary \Couleur{$\partial \Omega $}, but is also affected by source terms depending on the gravitational field, for which there is no univoque definition of the energy-momentum tensor.\\

However, according to Padmanabhan \cite{Padmanabhan2015}, there exists a suitable definition for what we will call an {\it energy current} four-vector \Couleur{$\Mat{\mathcal{G}}$} of the gravitational field, such that the total energy current \Couleur{$\Mat{\mathcal{P}}:=\Mat{\mathcal{G}}+{\bf P}$} is conserved:
\be\label{Conserv calP}
\Couleur{\mathcal{P}^\mu_{;\mu}=\mathcal{G}^\mu_{;\mu}+P^\mu_{;\mu}=0},
\ee
where the four-vector field \Couleur{${\bf P}$}, with components
\be\label{Matter current-GR}
\Couleur{P^\mu := +T^\mu _{\ \,\nu} v^\nu },
\ee
is the matter energy current associated with the matter distribution, for the observer(s) having the four-velocity field \Couleur{$\Mat{v}$}. \{The plus sign is with the $(+ - - -)$ signature that we use. This four-vector \Couleur{${\bf P}$} should not be confused with the three-vector \Couleur{$\Mat{P}$} in Eq. (\ref{Def P Sigma})$_1$. The gravitational current \Couleur{$\Mat{\mathcal{G}}$} of Ref. \cite{Padmanabhan2015} also depends on \Couleur{$\Mat{v}$}.\} Since, in coordinates adapted to the reference fluid defined by \Couleur{$\Mat{v}$}, we have  \cite{A44,Cattaneo1958}
\be\label{Vmu}
\Couleur{v^0 = \frac{1}{\sqrt{g_{00}}}, \qquad v^j=0},
\ee
we note that, in any such adapted coordinates:
\be
\Couleur{P^\mu=T^\mu _{\ \,0}/\sqrt{g_{00}}}.
\ee
In particular, in the Minkowski spacetime, and taking for \Couleur{$\Mat{v}$} the four-velocity field of some inertial reference frame, the gravitational current \Couleur{$\Mat{\mathcal{G}}$} vanishes, hence the conservation (\ref{Conserv calP}) of the (matter) current is exactly the same equation as the local energy conservation (\ref{Local Conserv E-2}) with the definitions (\ref{Def w Phi}). Whence our use of the expression ``matter energy current" to designate the four-vector field (\ref{Matter current-GR}). Thus, according to the result \cite{Padmanabhan2015}, the situation regarding the energy and momentum conservation would be nearly the same in GR as it is in the ether theory \cite{A35}, namely: (i) there is a (scalar) local conservation equation of the type (\ref{Local Conserv E}) for the total (material plus gravitational) energy; (ii) the energy density and its flux involved in that conservation equation depend on the reference frame;
\footnote{\ 
Actually, in the theory \cite{A35}, the energy density and its flux have been defined only in the preferred reference frame assumed by the theory.
}
(iii) there is no equivalent local conservation equation for the spatial momentum [by this we mean a (frame-dependent) space vector equation, thus three scalar equations, as Eq. (\ref{Local Conserv P})]. Anyway, as discussed in Ref. \cite{A35}, the conservation of the total momentum, when it takes place, precludes a conservation of the momentum of {\it matter}.

\subsection{Uniqueness and tensoriality of the Hilbert tensor}\label{Unique Hilbert tensor}

In addition to the difficulty described in the foregoing subsection, which did not seem solvable in the framework of the said theories (but may be solved by the work \cite{Padmanabhan2015}), there is a point that needs clarification. In a curved spacetime, the Hilbert tensor field \Couleur{$\Mat{T}$} is taken as the {\it source of the gravitational field} --- in general relativity and in many other relativistic theories of gravity. Clearly, that source has to be {\it locally} defined: it is not the global value (the space integral) of \Couleur{$\Mat{T}$} that matters to determine the gravitational field, but indeed the distribution of its local value. However, could not the Hilbert tensor be subject to ``relocalizations", due to the fact that the Lagrangian determining the equations of motion is not unique?\\

Let us add to the Lagrangian \Couleur{$L$} a total divergence: \Couleur{$L\hookrightarrow L'=L+D$} with
\be\label{add div to L}
\Couleur{D=\operatorname{div} \,\Mat{V} =\frac{1}{\sqrt{-g}} \partial _\rho  \left( V^\rho  \sqrt{-g}\right )},
\ee
where \Couleur{$V^\mu=V^\mu ({\bf q}^A, {\bf q}^A_{\mu },\underline{g}^{\mu \nu },\underline{g}^{\mu \nu }_{\rho })$} is a spacetime vector field. [Of course, the partial derivatives in Eq. (\ref{add div to L}) apply once the relevant fields have been substituted for the arguments of \Couleur{$V^\mu$}, see before Eq. (\ref{delta_S_coordinate-change-0}).] Then the Euler-Lagrange equations (\ref{E-L general}) stay {\it unchanged,} see e.g. Ref. \cite{Olver2012}. Note that, of course, the modified Lagrangian \Couleur{$L'$} is also an invariant scalar if \Couleur{$L$} is. But, {\it a priori}, should not the T-tensor generally change? This would indirectly contradict a statement of Forger \& R\"omer \cite{ForgerRoemer2004}, according to which the energy-momentum tensor field `` \Couleur{$\Mat{T}$} is the rank 2 tensor field on space-time \Couleur{$\mathrm{M}$} depending on the fields of the theory which satisfies 
\be\label{delta_S_metric-change-T}
\Couleur{\delta_g \int_{\mathrm{K}} \dd ^n {\bf X} \sqrt{\abs{\mathrm{det}\, g}}\,L =-\frac{1}{2}\int_{\mathrm{K}} \dd ^n {\bf X} \sqrt{\abs{\mathrm{det}\, g}}\, T^{\mu \nu }\,\delta g_{\mu \nu } }
\ee
for every compact subset \Couleur{$\mathrm{K}$} of \Couleur{$\mathrm{M}$} and for every variation \Couleur{$\delta g_{\mu \nu }$} of the metric tensor with support contained in \Couleur{$\mathrm{K}$} ". [Here, \Couleur{$n$} is the dimension of the spacetime, thus \Couleur{$n=4$} as far as we know. For us,  \Couleur{$\mathrm{K}$} is the compact closure of the bounded {\it open domain} \Couleur{$\mathrm{U}$}, \Couleur{$\mathrm{K} := \overline{\mathrm{U}} = \mathrm{U}\cup \partial \mathrm{U}$}; see Note \ref{K vs U}.] Indeed, from this statement one easily concludes that a total divergence does not change \Couleur{$T^{\mu \nu }$} \{see \cite{ForgerRoemer2004}, and see a detailed proof here around Eq. (\ref{S special-div})\}. However:\\

\noi ({\it a}) It is not precised what is meant exactly by a ``variation \Couleur{$\delta g_{\mu \nu }$} of the metric tensor" and what is meant exactly by \Couleur{$\delta _g$} (applied to the action integral) in the statement reproduced above, included in Theorem 4.2 of Ref. \cite{ForgerRoemer2004}. In the arguments (pp. 360--361) which lead the authors to state that theorem, the same situation is considered as in Subsect. \ref{Hilbert Tensor} hereabove. I.e., the variation of the metric occurs due to an infinitesimal coordinate change (or equivalently due to an infinitesimal diffeomorphism), generated by a {\it vector field} \Couleur{$\xi $} (noted \Couleur{$X $} in Ref. \cite{ForgerRoemer2004}), with support contained in \Couleur{$\mathrm{K}$}. (The variation of the metric is thus appropriately noted  \Couleur{$\delta _X g_{\mu \nu }$} in these arguments, and the variation of the action is noted \Couleur{$\delta _X S_{\mathrm{K}}$}.) It is in this precise situation that one can at the same time derive the basic equation (\ref{delta_S_coordinate-change}) and state that actually \Couleur{$\delta S_\mathrm{U}=0$} --- which is needed to derive the ``covariant conservation" (\ref{Semicolon_Cons}).\\

\noi ({\it b}) In that situation [i.e., the variation of the metric resulting thus from a coordinate change], Eq. (\ref{delta_S_metric-change-T}) is equivalent to our Eq. (\ref{delta_S_coordinate-change-T}). However, we do not see how it could be the case that the validity of Eq. (\ref{delta_S_coordinate-change-T}) ``for every [regular] compact subset \Couleur{$\mathrm{K}$} of \Couleur{$\mathrm{M}$} and for every variation \Couleur{$\delta g_{\mu \nu }$} of the metric tensor with support contained in \Couleur{$\mathrm{K}$}" would {\it characterize} (determine) some field object having components \Couleur{$T_{\mu \nu }$}. Indeed, since the action is invariant under coordinate changes, it follows that the l.h.s. of (\ref{delta_S_coordinate-change-T}) is automatically {\it zero } (for every [regular] compact subset \Couleur{$\mathrm{K}$} etc.). Hence, for example, \Couleur{$\underline{T_{\mu \nu }\equiv 0}$} {\it is a solution of (\ref{delta_S_coordinate-change-T}) (for every [regular] compact subset \Couleur{$\mathrm{K}$} etc.), as well as is (\ref{Hilbert tensor})}. \\

\noi ({\it c}) In that same situation, the following argument in Ref. \cite{ForgerRoemer2004} does not work. According to this argument, ``\Couleur{$\Mat{T}$} does not change when \Couleur{$L$} is modified by the addition of a total divergence (...), simply because the addition of such a term does not affect the l.h.s. of [Eq. (\ref{delta_S_metric-change-T})]." The last statement is true, but since in that situation the l.h.s. of Eq. (\ref{delta_S_metric-change-T}) is always zero, it can not prove that \Couleur{$\Mat{T}$} does not change. \\

\noi ({\it d}) Therefore, it seems that, instead of the foregoing situation, the ``variation \Couleur{$\delta g_{\mu \nu }$} of the metric tensor" alluded to in Theorem 4.2 of Ref. \cite{ForgerRoemer2004} be a variation of the metric itself (thus even in a fixed coordinate system):
\be\label{delta g per se}
\Couleur{ g^{\mu \nu }_{(\epsilon \Mat{h})}({\bf X}) =g^{\mu \nu }({\bf X})+\epsilon h ^{\mu \nu }({\bf X})},
\ee
where the field \Couleur{$\Mat{h}$}, with components \Couleur{$h^{\mu \nu }$}, is a given field of symmetric $(2\ 0)$ tensors defined on \Couleur{$\mathrm{U} $}. (The matter fields are thus left unchanged.) However, if that is indeed the case, then Eq. (\ref{delta_S_metric-change-T}) has a different meaning than Eq. (\ref{delta_S_coordinate-change-T}): for example, the variation \Couleur{$\delta g^{\mu \nu }$} of the metric now depends on the ten independent parameters \Couleur{$h ^{\mu \nu }\ (0\leq \mu \leq \nu \leq 3)$} instead of merely the four parameters \Couleur{$\xi ^\mu \ (\mu =0,...,3)$} as is the case in the arguments which lead to the statement of Theorem 4.2 in Ref. \cite{ForgerRoemer2004}, as well as in Subsect. \ref{Hilbert Tensor} hereabove. Thus, the validity of Eq. (\ref{delta_S_metric-change-T}) in that different situation has to be proved separately. This proof takes Points (i) and (ii) of the following theorem, whose conclusive part is its Point (iii).

\paragraph{Theorem 2.}\label{Theorem 2} {\it Let \Couleur{$L=L({\bf q}^A, {\bf q}^A_{\mu },\underline{g}^{\mu \nu },\underline{g}^{\mu \nu }_{\rho })$} be a Lagrangian that is defined and smooth whenever the determinant of the matrix \Couleur{$(\underline{g}^{\mu \nu })$} is negative, and that is invariant under general coordinate changes. Let \Couleur{$\mathrm{U}$} be a bounded open domain of the spacetime manifold \Couleur{$\mathrm{V}$} and let \Couleur{$\mathrm{K} = \overline{\mathrm{U}} = \mathrm{U} \cup \partial \mathrm{U} $} be its compact closure. Let \Couleur{$\mathrm{E}$} be the vector space of the symmetric \Couleur{$(2\ 0)$} tensor fields \Couleur{$\Mat{g}$} which are defined and continuous on \Couleur{$\mathrm{K}$} and which are \Couleur{$\mathcal{C}^1$} on \Couleur{$\mathrm{U}$}. Let \Couleur{$\mathrm{E}^\star $} be the subset of \Couleur{$\mathrm{E}$} made of the tensor fields \Couleur{$\Mat{g}\in \mathrm{E}$} such that, for any chart \Couleur{$\chi $} defined in a neighborhood of \Couleur{$\mathrm{K}$}} (assuming there does exist such charts), {\it we have \Couleur{$g^{-1}:=\mathrm{det}\,G<0$} over the domain \Couleur{$ \chi (\mathrm{K})$}, where \Couleur{$G$} is the component matrix \Couleur{$G=(g^{\mu \nu })$}. The matter fields \Couleur{$\phi ^A \ (A=1,...,n)$} being given functions which are defined and continuous on \Couleur{$\mathrm{K}$} and which are \Couleur{$\mathcal{C}^1$} on \Couleur{$\mathrm{U}$}, define an invariant functional \Couleur{$S$} on \Couleur{$\mathrm{E}^\star $} by}
\be\label{S special}
\Couleur{S(\Mat{g}):= 
\int_{\Omega } \mathcal{L}(\Mat{\phi} ^A({\bf X}),\Mat{\phi} ^A_{,\mu  }({\bf X}),g^{\mu \nu}({\bf X}),g^{\mu \nu}_{,\rho} ({\bf X})) \, \dd ^4 {\bf X} ,\quad \Omega :=\chi (\mathrm{U}),\ \mathcal{L}:=\sqrt{-g}L}.
\ee

\noi (i) {\it Given any two tensor fields  \Couleur{$\Mat{g}\in \mathrm{E}^\star $} and \Couleur{$\Mat{h}\in \mathrm{E}$}, there is a number \Couleur{$a=a(\Mat{g},\Mat{h})>0$} such that, for \Couleur{$\epsilon \in ]-a,+a[$}, the tensor field \Couleur{$\Mat{g}+\epsilon \Mat{h}$} is in \Couleur{$\mathrm{E}^\star $}. We have:}  
\be\label{delta S special}
\Couleur{ \Delta _{\Mat{g}\,\Mat{h}}\,S:=\left(\frac{\dd }{\dd \epsilon }S(\Mat{g}+\epsilon \Mat{h})\right)_{\epsilon =0}=\int_{\Omega } \left [\frac{\partial \mathcal{L}}{\partial \underline{g}^{\mu \nu }} h^{\mu \nu }+ \frac{\partial \mathcal{L}}{\partial \left(\underline{g}^{\mu \nu }_{\rho }\right)} h^{\mu \nu }_{ ,\rho }\right ]_\Mat{g} \, \dd ^4 {\bf X}},\qquad \qquad
\ee
{\it where the subscript \Couleur{$_\Mat{g}$} means that, at any \Couleur{${\bf X}\in \Omega $}, one considers the derivatives of the function  \Couleur{$\mathcal{L}$} for the values \Couleur{$\underline{g}^{\mu \nu}=g^{\mu \nu}({\bf X}),\ \underline{g}^{\mu \nu}_\rho = g^{\mu \nu}_{,\rho} ({\bf X})$} of its arguments.}\\

\noi (ii) {\it If \Couleur{$\Mat{g}\in \mathrm{E}^\star $}, \Couleur{$\Mat{h}\in \mathrm{E}$}, and \Couleur{$\Mat{h}(X)=0$} when \Couleur{$X \in \partial \mathrm{U}$}, then we have, in any chart \Couleur{$\chi $} whose domain contains \Couleur{$\mathrm{K}$}:} 
\be\label{delta S special-2}
\Couleur{ \Delta _{\Mat{g}\,\Mat{h}}\,S = \frac{1}{2} \int_{\Omega } T_{\mu \nu }\,h^{\mu \nu }\, \sqrt{-g}\,\dd ^4 {\bf X}},
\ee
{\it where the \Couleur{$T_{\mu \nu }$} 's are defined in Eq. (\ref{Hilbert tensor}), the derivatives being taken as in the subscript notation \Couleur{$_\Mat{g}$} above.}\\

\noi (iii) {\it Equation (\ref{delta S special-2}) determines uniquely the continuous functions \Couleur{${\bf X} \mapsto T_{\mu \nu }({\bf X})$}, \Couleur{$\chi (\mathrm{K}) \rightarrow \mathbb{R}$}. It follows that the ``Hilbert tensor" with components (\ref{Hilbert tensor}) is not modified by the addition of a four-divergence. Moreover, this is indeed a \Couleur{$(0\ 2)$} tensor field. } \\

\noi {\it Proof.} As a preliminary, recall that, if the matrix \Couleur{$G=(g^{\mu \nu })$} of some \Couleur{$(2\ 0)$} tensor field \Couleur{$\Mat{g}$} in one chart \Couleur{$\chi $} verifies \Couleur{$g^{-1}:=\mathrm{det}\,G<0$} over \Couleur{$\chi (\mathrm{K})$}, then the corresponding matrix \Couleur{$G'$} in any other chart \Couleur{$\chi '$} in the atlas of \Couleur{$\mathrm{V}$}, whose domain also contains \Couleur{$\mathrm{K}$}, verifies \Couleur{$g'^{-1}:=\mathrm{det}\,G'<0$} over \Couleur{$\chi' (\mathrm{K})$}: indeed the Jacobian matrix \Couleur{$J=(\frac{\partial x'^\mu }{\partial x^\nu })$} is invertible for these two compatible charts, and we have \Couleur{$g'^{-1}=g^{-1}\,(\mathrm{det}\,J)^2$}. Also remind that the invariance of the functional (\ref{S special}) under the change of the chart follows from the definition of a Lagrangian that is invariant under general coordinate changes, Eq. (\ref{L invariant}), and from the invariance of the four-volume measure \Couleur{$\dd V_4 :=\sqrt{-g}\dd ^4 {\bf X}$}.\\

(i) Let \Couleur{$\Mat{g}\in \mathrm{E}^\star $} and \Couleur{$\Mat{h}\in \mathrm{E}$}, thus in particular these are two continuous functions defined over the compact set \Couleur{$\mathrm{K}$}. The real function \Couleur{$M \mapsto \phi (M):=\mathrm{det}\,M$} is defined and \Couleur{$\mathcal{C}^1$} over the vector space \Couleur{${\sf M}(4,\mathbb{R})$} of the real \Couleur{$4\times 4$} matrices. Choose a chart \Couleur{$\chi $} defined in a neighborhood of \Couleur{$\mathrm{K}$}. For \Couleur{${\bf X} \in \chi (\mathrm{K})$}, we note \Couleur{$G({\bf X})$} the matrix \Couleur{$(g^{\mu \nu }({\bf X}))$}. Let us note also \Couleur{$H({\bf X})=(h^{\mu \nu }({\bf X}))$}. Since \Couleur{$\Mat{g}\in \mathrm{E}^\star $}, we have \Couleur{$\phi (G({\bf X})) < 0$} for \Couleur{${\bf X} \in \chi (\mathrm{K})$}. Because \Couleur{$\phi \circ G$} is a continuous function over the compact \Couleur{$ \chi (\mathrm{K})$}, it is bounded and reaches its bounds. Hence, for some number \Couleur{$d<0$}, we have \Couleur{$\phi (G({\bf X}))\leq  d$} for \Couleur{${\bf X} \in \chi (\mathrm{K})$}. 
The set of matrices \Couleur{$\mathrm{C}=\{M=G({\bf X})+\epsilon H({\bf X});{\bf X}\in \chi (\mathrm{K})\ \mathrm{and}\ \abs{\epsilon }\leq 1\}$} is compact, hence we have \Couleur{$\mathrm{Sup} \{\parallel \phi '(M)\parallel ;\,M\in \mathrm{C}\}=A<\infty $}. Also, \Couleur{$\mathrm{Sup} \{\parallel H({\bf X})\parallel ;\,{\bf X}\in \chi (\mathrm{K})\}=B<\infty $}. Therefore, we have for any \Couleur{${\bf X}\in \chi (\mathrm{K})$} and any \Couleur{$\epsilon \in ]-1,1[$}:
\be
\Couleur{\abs{\phi (G({\bf X})+\epsilon H({\bf X}) )-\phi (G({\bf X}))} \leq A \abs{\epsilon} \parallel H({\bf X})\parallel \leq \abs{\epsilon}  A B}.
\ee
Hence, there is a number \Couleur{$a>0$} such that, for \Couleur{$\epsilon \in ]-a,+a[$}, we have \Couleur{$\phi (G({\bf X})+\epsilon H({\bf X}) )< d/2 <0$} for any \Couleur{${\bf X}\in \chi (\mathrm{K})$}. Thus, for \Couleur{$\epsilon \in ]-a,+a[$}, we have \Couleur{$\Mat{g}+\epsilon\Mat{h}\in \mathrm{E}^\star $}, as announced. Since that statement does not depend on the chart, the number \Couleur{$a>0$} does not depend on the chart, either, thus \Couleur{$a=a(\Mat{g},\Mat{h})>0$}.\\

Denoting henceforth \Couleur{$\Mat{g}_{(\epsilon \Mat{h})}:=\Mat{g}+\epsilon \Mat{h}$} for brevity, we define a \Couleur{$\mathcal{C}^1$} function \Couleur{$f$} from \Couleur{$\Omega \times ]-a,+a[$} into \Couleur{$\mathbb{R}$} by setting 
\be\label{Def f}
\Couleur{f({\bf X},\epsilon )= \mathcal{L}(\Mat{\phi} ^A({\bf X}),\Mat{\phi} ^A_{,\mu  }({\bf X}), g_{(\epsilon \Mat{h})}^{\mu \nu}({\bf X}),\partial _\rho g_{(\epsilon \Mat{h})}^{\mu \nu} ({\bf X}))}.
\ee
Using the definition (\ref{delta g per se}), we have:
\be
\Couleur{\frac{\partial \left( g^{\mu \nu }_{(\epsilon \Mat{h})}({\bf X})\right)}{\partial \epsilon } = h^{\mu \nu }({\bf X})}, \quad  \Couleur{\frac{\partial \left( \partial _\rho g^{\mu \nu }_{(\epsilon \Mat{h})}({\bf X})\right)}{\partial \epsilon } = \frac{\partial }{\partial \epsilon }\left(g^{\mu \nu }_{,\rho }({\bf X})+\epsilon h^{\mu \nu }_{,\rho }({\bf X}) \right )=h^{\mu \nu }_{,\rho }({\bf X})}.
\ee
Therefore, we get
\bea\label{df/d eps}\nonumber
\Couleur{\frac{\partial f}{\partial \epsilon }\left ({\bf X},\epsilon \right)} & = & \Couleur{\left(\frac{\partial \mathcal{L}}{\partial \underline{g}^{\mu \nu }} \right)_{g_{(\epsilon \Mat{h})}} \frac{\partial \left( g^{\mu \nu }_{(\epsilon \Mat{h})}({\bf X})\right)}{\partial \epsilon } + \left ( \frac{\partial \mathcal{L}}{\partial \left(\underline{g}^{\mu \nu }_{\rho }\right)} \right)_{g_{(\epsilon \Mat{h})}}   \frac{\partial \left( \partial _\rho g^{\mu \nu }_{(\epsilon \Mat{h})}({\bf X})\right)}{\partial \epsilon }}\\
& = & \Couleur{\left(\frac{\partial \mathcal{L}}{\partial \underline{g}^{\mu \nu }} \right)_{g_{(\epsilon \Mat{h})}} h^{\mu \nu }({\bf X}) + \left ( \frac{\partial \mathcal{L}}{\partial \left(\underline{g}^{\mu \nu }_{\rho }\right)} \right)_{g_{(\epsilon \Mat{h})}}  h^{\mu \nu }_{,\rho }({\bf X})}. 
\eea 
From the definitions (\ref{S special}) and (\ref{Def f}), we have \Couleur{$S(\Mat{g}_{(\epsilon \Mat{h})})= \int_{\Omega } f ({\bf X},\epsilon ) \, \dd ^4 {\bf X} $},  hence 
\be\label{d S/d eps}
\Couleur{ \frac{\dd }{\dd \epsilon }S(\Mat{g}_{(\epsilon \Mat{h})}) =\int_\Omega \frac{\partial f}{\partial \epsilon }\left ({\bf X},\epsilon \right) \, \dd ^4 {\bf X}}.
\ee
Since at \Couleur{$\epsilon =0$} we have \Couleur{$\Mat{g}_{(\epsilon \Mat{h})}=\Mat{g}$}, Eq. (\ref{delta S special}) follows from Eqs. (\ref{df/d eps}) and (\ref{d S/d eps}).\\

\noi (ii) The second term in the integrand on the r.h.s. of (\ref{delta S special}) can be written as
\bea\label{surface term}
\Couleur{\left (\frac{\partial \mathcal{L}}{\partial \left(\underline{g}^{\mu \nu }_{\rho }\right)} \right )_\Mat{g} h^{\mu \nu }_{ ,\rho }({\bf X})} & = &
\Couleur{\frac{\partial }{\partial x^\rho } \left( 
\left (\frac{\partial \mathcal{L}}{\partial \left(\underline{g}^{\mu \nu }_{\rho }\right)}\right )_\Mat{g}
 h^{\mu \nu }({\bf X})\right )} \\
& & \Couleur{- h^{\mu \nu }({\bf X}) \frac{\partial }{\partial x^\rho } \left( 
\frac{\partial \mathcal{L}}{\partial \left(\underline{g}^{\mu \nu }_{\rho }\right)}
\right )_\Mat{g}}.\nonumber
\eea
The first term on the r.h.s. of (\ref{surface term}) is a divergence in \Couleur{$\mathbb{R}^4$} and its integral on \Couleur{$\Omega$} vanishes if \Couleur{$h^{\mu \nu }({\bf X})=0 \ (\mu ,\nu =0,...,3)$} on \Couleur{$\partial \Omega$}, thus if \Couleur{$\Mat{h}(X)=0$} on \Couleur{$\partial \mathrm{U}$}. [Remind: \Couleur{$\partial \Omega = \partial\, (\chi (\mathrm{U}))=\chi (\partial \mathrm{U}) $}.] Hence, in that case, (\ref{delta S special}) rewrites as 
\be\label{delta S special-3}
\Couleur{ \Delta _{\Mat{g}\,\Mat{h}}\,S = \int_{\Omega } \left [\frac{\partial \mathcal{L}}{\partial \underline{g}^{\mu \nu }} - \frac {\partial }{\partial x^\rho } \left (\frac{\partial \mathcal{L}}{\partial \left(\underline{g}^{\mu \nu }_{\rho }\right)} \right ) \right ]_\Mat{g} \, h^{\mu \nu } \dd ^4 {\bf X}}.
\ee
In view of Eq. (\ref{Hilbert tensor}), this is Eq. (\ref{delta S special-2}).\\

\noi (iii) Consider a given tensor field \Couleur{$\Mat{g}\in \mathrm{E}^\star $} and, in a given chart \Couleur{$\chi $} whose domain contains \Couleur{$\mathrm{K}$}, let \Couleur{${\bf X} \mapsto T_{\mu \nu }({\bf X})$} and \Couleur{${\bf X} \mapsto \widetilde{T}_{\mu \nu }({\bf X})$} be two sets of functions (\Couleur{$\mu,\nu =0,..., 3$}) defined and continuous over \Couleur{$\chi (\mathrm{K})$}, each set being symmetric, such that both verify Eq. (\ref{delta S special-2}) for any tensor field \Couleur{$\Mat{h}\in \mathrm{E}$} that vanishes on \Couleur{$ \partial \mathrm{U}$}. [We do {\it not} assume that either \Couleur{$T_{\mu \nu }$} or \Couleur{$\widetilde{T}_{\mu \nu }$} is given by Eq. (\ref{Hilbert tensor}).] We claim that \Couleur{$T_{\mu \nu }= \widetilde{T}_{\mu \nu }$} over \Couleur{$\chi (\mathrm{K}) $}. Denoting \Couleur{$\delta T_{\mu \nu }:=\widetilde{T}_{\mu \nu }-T_{\mu \nu }$}, we thus have for any such tensor field \Couleur{$\Mat{h}$}:
\be\label{integral T=0}
\Couleur{\int_\Omega  \delta T_{\mu \nu }\,h^{\mu \nu }\, \sqrt{-g}\,\dd ^4 {\bf X}=0}.
\ee
Consider a given pair \Couleur{$(\mu _0,\nu _0)$} of indices. Let \Couleur{$\varphi $} be any real function which is defined and continuous over \Couleur{$\chi (\mathrm{K}) $}, which is \Couleur{$\mathcal{C}^1 $} over \Couleur{$\Omega$}, and that has compact support \Couleur{$\mathrm{K}'\subset \Omega $}. 
Hence, \Couleur{$\varphi ({\bf X}) \ne 0$} implies \Couleur{${\bf X} \in \Omega $}. If \Couleur{${\bf X} \in \partial \Omega $}, we have \Couleur{${\bf X} \notin \Omega$} since \Couleur{$\Omega $} is open, hence \Couleur{$ \varphi ({\bf X}) = 0 $}. Therefore, by setting \Couleur{$h^{\mu \nu }=\frac{1}{2}(\delta ^\mu _{\mu _0} \delta ^\nu _{\nu _0}+\delta ^\mu _{\nu _0} \delta ^\nu _{\mu _0})\varphi $} in the chart \Couleur{$\chi $}, we define a tensor field \Couleur{$\Mat{h}\in \mathrm{E}$} such that \Couleur{$\Mat{h}(X)=0$} for \Couleur{$X\in \partial \mathrm{U} $}. We can thus apply (\ref{integral T=0}) to get 
\be\label{integral T=0 fixed indices}
\Couleur{\int_\Omega  (\delta T_{\mu_0 \nu_0 }+\delta T_{\nu_0 \mu_0 })\, \sqrt{-g}\,\varphi  \,\dd ^4 {\bf X}}= \Couleur{\int_\Omega  (2\delta T_{\mu_0 \nu_0 }\, \sqrt{-g})\,\varphi  \,\dd ^4 {\bf X}=0}.
\ee
(The second equality follows from the symmetry of \Couleur{$T_{\mu \nu }$} and \Couleur{$\widetilde{T}_{\mu \nu }$}.) Since this is true for any such function \Couleur{$\varphi $}, we deduce that \Couleur{$\delta T_{\mu_0 \nu_0 }\, \sqrt{-g}$} is zero almost everywhere in \Couleur{$\Omega $}. But since this is a continuous function, it is zero everywhere in the open set \Couleur{$\Omega \subset \mathbb{R}^4$}, and therefore it is zero also in its closure \Couleur{$\overline{\chi (\mathrm{U})}=\chi(\overline{\mathrm{U}}) =\chi (\mathrm{K})$}. Then, since \Couleur{$\sqrt{-g}\ne 0$} over \Couleur{$\chi (\mathrm{K})$}, we have \Couleur{$\delta T_{\mu_0 \nu_0 }=0$} over \Couleur{$\chi (\mathrm{K})$}. This proves our precise statement about uniqueness at the beginning of this paragraph.\\

Now suppose the Lagrangian is a four-divergence: \Couleur{$L=((\sqrt{-g}\,V^\mu )_{,\mu })/\sqrt{-g}$} with \Couleur{$V^\mu=V^\mu ({\bf q}^A, {\bf q}^A_{\mu },\underline{g}^{\mu \nu },\underline{g}^{\mu \nu }_{\rho })$}. Then the integral (\ref{S special}) rewrites as:
\be\label{S special-div}
\Couleur{S(\Mat{g}+\epsilon \Mat{h})= \int_{\partial \Omega } U^\mu(\Mat{\phi} ^A({\bf X}),\Mat{\phi} ^A_{,\mu  }({\bf X}),g_{(\epsilon \Mat{h})}^{\mu \nu}({\bf X}),(g_{(\epsilon \Mat{h})}^{\mu \nu})_{,\rho} ({\bf X})) \dd S_\mu({\bf X})}
\ee
(setting \Couleur{$U^\mu :=\sqrt{-g}\,V^\mu$}, and with \Couleur{$\Mat{g}_{(\epsilon \Mat{h})}:=\Mat{g}+\epsilon \Mat{h}$} and \Couleur{$\dd S_\mu :=\varepsilon _{\mu \nu \rho \sigma } \dd x^\nu \wedge \dd x^\rho \wedge \dd x^\sigma $}) --- when this integral makes sense, which is true if \Couleur{$\Mat{g}\in \mathrm{E}^\star $}, \Couleur{$\Mat{h}\in \mathrm{E}$},  and \Couleur{$\abs{\epsilon}<a=a(\Mat{g},\Mat{h})$}. If, moreover, \Couleur{$\Mat{h}(X)=0$} on \Couleur{$\partial \mathrm{U}$}, we have \Couleur{$\Mat{g}_{(\epsilon \Mat{h})}(X)=\Mat{g}(X)$} for any \Couleur{$X \in \partial \mathrm{U} $}, so that the integral (\ref{S special-div}) does not depend on \Couleur{$\epsilon \in ]-a,a[$}. Therefore, the l.h.s. of Eq. (\ref{delta S special-2}) is zero. Since we have shown that this equation determines uniquely the functions \Couleur{${\bf X} \mapsto T_{\mu \nu }({\bf X}), \chi (\mathrm{K}) \rightarrow \mathbb{R}$}, these functions are zero. \\

Let us finally prove the actual tensoriality of the ``Hilbert tensor", whose components are defined by Eq. (\ref{Hilbert tensor}). Considering now any two charts \Couleur{$\chi $} and \Couleur{$\chi' $} whose domain contains \Couleur{$\mathrm{K}$}, Eq. (\ref{delta S special-2}) is true for any tensor field \Couleur{$\Mat{h}\in \mathrm{E}$} that vanishes on \Couleur{$ \partial \mathrm{U}$}, using either \Couleur{$\chi $} or \Couleur{$\chi' $} on the r.h.s. (with primes for \Couleur{$\chi' $}). Since the l.h.s. of (\ref{delta S special-2}) is invariant as is the action, so is the r.h.s.; i.e., we have for any such \Couleur{$\Mat{h}$}:
\be\label{invariance delta S}
\Couleur{ \int_{\Omega } T_{\mu \nu }\,h^{\mu \nu }\, \sqrt{-g}\,\dd ^4 {\bf X} = \int_{\Omega' } T'_{\mu \nu }\,h'^{\mu \nu }\, \sqrt{-g'}\,\dd ^4 {\bf X}'}.
\ee
Composing with the reverse coordinate maps \Couleur{$\chi^{-1} $} and \Couleur{$\chi'^{-1} $}, we may regard \Couleur{$ T_{\mu \nu }, h^{\mu \nu }, T'_{\mu \nu }, h'^{\mu \nu }$} as functions defined over \Couleur{$\mathrm{K}=\chi^{-1}(\chi (\mathrm{K}) )= \chi'^{-1}(\chi' (\mathrm{K}))  $}, and we have for any such \Couleur{$\Mat{h}$}:
\be\label{invariance delta S-U}
\Couleur{ \int_{\mathrm{U} } T_{\mu \nu }\,h^{\mu \nu }\, \dd V_4= \int_{\mathrm{U} } T'_{\mu \nu }\,h'^{\mu \nu }\, \dd V_4}.
\ee
For any given tensor field \Couleur{$\Mat{h}_0\in \mathrm{E}$} 
, set \Couleur{$f:=T_{\mu \nu }\,h_{0}^{\mu \nu }$} and \Couleur{$f':=T'_{\mu \nu }\,h'^{\mu \nu }_{0}$}, which are thus two continuous functions on \Couleur{$\mathrm{K}$}. Consider any open domain \Couleur{$\mathrm{W}$} with \Couleur{$\mathrm{W}\subset \mathrm{U}$}. Take any function \Couleur{$\phi $} which is defined and continuous over \Couleur{$\mathrm{K} $}, which is \Couleur{$\mathcal{C}^1 $} over \Couleur{$\mathrm{U}$}, and that has compact support \Couleur{$\mathrm{K}'\subset \mathrm{W} $}.  Define \Couleur{$\Mat{h}:=\phi \,\Mat{h}_0$}. This is a tensor field that belongs to \Couleur{$\mathrm{E}$} and vanishes on \Couleur{$ \partial \mathrm{W}$}. Therefore, the open domain \Couleur{$\mathrm{U}$} being arbitrary in the already proved point (ii) and hence in (\ref{invariance delta S-U}), we can apply (\ref{invariance delta S-U}) with \Couleur{$\mathrm{W}$} instead of \Couleur{$\mathrm{U}$}. We thus get:
\be\label{invariance delta S-W}
\Couleur{ \int_{\mathrm{W} } f\phi \, \dd V_4= \int_{\mathrm{W} } f'\phi \, \dd V_4}.
\ee
Because this is true for any such function \Couleur{$\phi $}, it follows that we have \Couleur{$f=f'$} almost everywhere in \Couleur{$\mathrm{W}$}. Since these are continuous functions, we have \Couleur{$f=f'$} in \Couleur{$\mathrm{W} $}. And since this is true for any open domain \Couleur{$\mathrm{W}$} with \Couleur{$\mathrm{W}\subset \mathrm{U}$}, we have \Couleur{$f=f'$}  in \Couleur{$\mathrm{U} $}, and hence also in \Couleur{$\mathrm{K}=\overline{ \mathrm{U}}$}. That is, \Couleur{$T_{\mu \nu }\,h_{0}^{\mu \nu }$} is invariant under coordinate changes, for whatever tensor field \Couleur{$\Mat{h}_0\in \mathrm{E}$}. Considering a given point \Couleur{$X\in \mathrm{K}$}, we define a linear form \Couleur{$\Phi $} on the vector space \Couleur{$\mathcal{T}^2_0$} of the \Couleur{$(2\ 0)$} tensors at \Couleur{$X$}, by setting:
\be\label{Def Phi}
\Couleur{\forall \underline{\Mat{h}}\in \mathcal{T}^2_0,\ \Phi (\underline{\Mat{h}}):=T_{\mu \nu }(X)\underline{h}^{\mu \nu }},
\ee
which is thus independent of the chart. But the dual space of \Couleur{$\mathcal{T}^2_0$} is known (and easily checked) to be the vector space \Couleur{$\mathcal{T}^0_2$} of the \Couleur{$(0\ 2)$} tensors \Couleur{$\underline{\Mat{S}}$} at \Couleur{$X$}. Hence, there is a unique tensor \Couleur{$\underline{\Mat{S}}\in \mathcal{T}^0_2$} for which, in any chart, we have
\be\label{Phi=Phi_S}
\Couleur{\forall \underline{\Mat{h}}\in \mathcal{T}^2_0,\ \Phi (\underline{\Mat{h}})=S_{\mu \nu }\underline{h}^{\mu \nu }}.
\ee
From (\ref{Def Phi}) and (\ref{Phi=Phi_S}), it follows that, in any chart, the numbers \Couleur{$T_{\mu \nu }(X)$} are the components of the unique tensor \Couleur{$\underline{\Mat{S}}\in \mathcal{T}^0_2$}, Q.E.D. \hfill $\square$ \\

\section{A uniqueness result for the energy balance}\label{Uniqueness E}
\subsection{Is the energy balance equation unique?}

We begin with a discussion of this question for a system of isentropically deformable media in Newtonian gravity (NG). The energy balance (\ref{Newton Matter Energy Balance}) established in Section \ref{NG} for the {\it matter field equations} of NG has the form
\be\label{Matter Balance}
\Couleur{\partial _\mu V^\mu = \operatorname{field\ source} := -\rho \frac{\partial U}{\partial t}},
\ee
with the four-components  column vector \Couleur{$(V^\mu)$} being here the ``matter current" made with the matter energy density and flux: 
\be\label{V^mu m}
\Couleur{(V^\mu)=(w_{\operatorname{m}},\Mat{\Phi }_{\operatorname{m}}) }. 
\ee
As we saw, Eq. (\ref{Matter Balance}) [i.e. Eq. (\ref{Newton Matter Energy Balance})] is verified as soon as the following three equations are verified among the matter field equations: Newton's second law (\ref{Newton continuum}), the isentropy equation (\ref{Isentropy}), and the continuity equation (\ref{Continuity Eqn}). For instance, we did not use the ``constitutive equation" that relates the stress tensor to some deformation tensor, or (for a barotropic fluid) that relates the pressure with the density. We note that, in view of Eqs. (\ref{w_m NG}) and (\ref{Phi_m NG}), the matter current \Couleur{$(V^\mu)$} is polynomial in the local values of the fields \Couleur{$\phi ^A \ (A=1,...,n)=(\rho ,{\bf v},\Pi ,\Mat{\sigma },U)$} that appear in those equations. (Thus, {\it assigning in this section --- contrary to Section \ref{T-tensor} --- a different number \Couleur{$A$} to different components of a given vector or tensor field:} \Couleur{$n=12$} here. The gravitational potential \Couleur{$U$} plays the same role as does the metric tensor in a Lagrangian for the matter fields in a curved spacetime, as was the case in the foregoing section.) Now we ask if we can find a different expression for the matter current, say \Couleur{$V'^\mu$}, for which the l.h.s. of (\ref{Matter Balance}) would be {\it always} the same as with the current (\ref{V^mu m}), so that the same balance equation (\ref{Matter Balance}) would be valid with \Couleur{$V'^\mu$}, when it is with \Couleur{$V^\mu$}.Thus: Can we change the matter current \Couleur{$V^\mu$} for another one \Couleur{$V'^\mu=V^\mu +W^\mu$}, also polynomial with respect to the local values of the fields at any spacetime point \Couleur{$X$}, \Couleur{$q ^A=\phi ^A(X)$}, so that the l.h.s. of (\ref{Matter Balance}) would be unchanged {\it for whatever values of the fields?} I.e., can we find a column four-vector \Couleur{$W^\mu$} which would be polynomial in the \Couleur{$q ^A$} 's, and such that we would have \Couleur{$\partial _\mu W^\mu\equiv 0$}?



\subsection{A uniqueness result}

Thus, let \Couleur{$W^\mu$} be an order-\Couleur{$N$} polynomial in the field values \Couleur{$q ^A=\phi ^A(X)$}, its coefficients being allowed to depend on the spacetime position \Couleur{$X$}:
\be\label{W mu}
\Couleur{W^\mu(X,q ^A)=C^\mu _0(X) + C^\mu _{1\, A}(X)\,q ^A+...+C^\mu _{N\, A_1...A_N}(X)\,q ^{A_1}...q ^{A_N}}\quad (\Couleur{A_1\leq ...\leq A_N}).
\ee
Assume that its 4-divergence vanishes identically, \Couleur{$\partial _\mu W^\mu\equiv 0$}:
\bea\label{Div W mu}
\Couleur{0} & \equiv & \Couleur{C^\mu _{0\,,\mu } + C^\mu _{1\,A ,\mu }q ^A + C^\mu _{1\, A}q ^A_{\mu }+ ...+C^\mu _{N\, A_1...A_N\, ,\mu }q ^{A_1}...q ^{A_N} } \\
& & \Couleur{+ C^\mu _{N\, A_1...A_N}\left(q ^{A_1}_{\mu }q ^{A_2}...q ^{A_N}+...+q ^{A_1}...q ^{A_{N-1}}q ^{A_N}_{\mu }\right )} \quad (\Couleur{q^A_\mu :=\phi ^A_{,\,\mu }(X)}). \nonumber
\eea
I.e., at any spacetime point \Couleur{$X_0(x^\rho) $}, Eq. (\ref{Div W mu}) is valid for whatever {\it possible} values \Couleur{$q^A$} and  \Couleur{$q ^A_\mu $} of the fields and their derivatives at \Couleur{$X_0$}. But, for {\it whatever} values \Couleur{$q^A$} and  \Couleur{$q ^A_\mu $} of these variables, there exists smooth functions \Couleur{$X\mapsto \phi  ^A(X)\ (A=1,...,n)$}, defined in some neighborhood \Couleur{$\mathrm{U}$} of \Couleur{$X_0$}, such that we have 
\be
\Couleur{\phi  ^A(X_0)=q^A\ (A=1,...,n)},\ \mathrm{and}\ \Couleur{\frac{\partial \phi  ^A}{\partial x^\mu }(X_0)= q^A_\mu\ (A=1,...,n;\ \mu =0,...,3)}.
\ee
Thus our assumption means that on the r.h.s. of (\ref{Div W mu}) the polynomial function in the real variables \Couleur{$q^A$} and  \Couleur{$q ^A_\mu $} \Couleur{$(A=1,...,n;\ \mu =0,...,3)$} 
is identically zero. Hence its coefficients are all zero. In particular:
\be\label{C=0}
\Couleur{C^\mu _{1\, A}(X_0)=0,..., C^\mu _{N\, A_1...A_N}(X_0)=0}.
\ee
Thus all coefficients in (\ref{W mu}) are zero --- except perhaps \Couleur{$C^\mu _0$}, with \Couleur{$C^\mu _{0,\,\mu}=0 $}.\\

We thus got that we {\it cannot} alter the analytical expression of \Couleur{$w_{\operatorname{m}}$} and \Couleur{$\Mat{\Phi }_{\operatorname{m}} $} on the l.h.s. of the matter energy balance (\ref{Newton Matter Energy Balance}). [Apart from arbitrarily adding a zero-divergence vector field \Couleur{$C^\mu _0$} that is {\it independent} of the matter fields --- this is indeed obviously possible, but we can get rid of this by asking that the matter current \Couleur{$(V^\mu)$} be polynomial in the fields {\it and have no zero-order term,} as is indeed the case in all concrete examples.] The gravitational energy balance (\ref{Newton Grav Energy Balance}) has just the same form:
\be\label{Gravitational Balance}
\Couleur{\partial _\mu V^\mu = \operatorname{matter\ source} := \rho \frac{\partial U}{\partial t}},
\ee
where \Couleur{$V^\mu=(w_{\operatorname{g}},\Mat{\Phi }_{\operatorname{g}})$} is polynomial in the gravitational field \Couleur{$\phi ^A \ (A=1,...,4)=(\partial _\mu U)$}. It is valid when the gravitational field equation is. Therefore, similarly as we found for the matter field energy balance, we cannot alter the analytical expression (\ref{Newton Grav Energy Balance}) of the gravitational energy balance.



\subsection{Generalization}
These results are clearly general. Consider e.g. the Maxwell electromagnetic field instead of the Newtonian gravitational field. The energy balance of the e.m. field is:
\be\label{Maxwell Field Energy}
\Couleur{ \frac{\partial w_{\operatorname{em} }}{\partial  t} +\operatorname{div} \,\Mat{\Phi }_{\operatorname{em}}=-{\bf j.E}},
\ee
with \Couleur{$w_{\operatorname{em}}:= \frac{   {\bf E}^2+{\bf B}^2}{8\pi}$} the {\it volume energy density of the electromagnetic field,} and \Couleur{$\Mat{\Phi }_{\operatorname{em}} := \frac{{\bf E}\wedge {\bf B}}{4\pi }$}  the {\it electromagnetic energy flux}. The same uniqueness result says that we cannot find an alternative expression for \Couleur{$w_{\operatorname{em}}$} and \Couleur{$\Mat{\Phi }_{\operatorname{em}} $} on the l.h.s., which would be valid for whatever values of the fields \Couleur{${\bf E}$} and \Couleur{${\bf B}$} and their first derivatives.

\section{Conclusion}

The classical concept of energy emerges from an analysis of the power done, first in the case of a mass point and then for the case of a volume element in a continuous medium. We have argued that, in the case of a continuous medium or a system of fields, the meaning of the energy conservation is primarily local: it says that, in {\it any} bounded domain, the energy loss or gain is due only to a well-identified flux that goes through the boundary of that domain. Thus it expresses in a general way the Lavoisier principle: ``Nothing is lost, nothing is created, everything transforms." While it is of course interesting also and even often important to have global energy conservation laws, this interest is limited by two facts: i) an exact global conservation law can be hoped, strictly speaking, only for the Universe as a whole, because there are energy exchanges at all scales --- but physics can not be reduced to cosmology. ii) A global energy conservation law says merely that one number is a constant: the total energy; in the most favorable case with global conservation of the energy-momentum and angular momentum, ten numbers are constant. In the relevant case of a system of fields, however, there is an infinite number of degrees of freedom, so this is only a small part of the information needed.\\

We have tried to precisely state and prove the main results regarding the derivation of the Hilbert tensor from the invariance of the action in generally-covariant theories. We hope to have proved these results in a convincing way, keeping the mathematical sophistication to the minimum needed. The Hilbert tensor's theory is beautiful and is essential to general relativity. It is important also in relativistic quantum mechanics. One should note, however, that historically the main examples of the energy-momentum tensor have been derived from the corresponding local conservation equations for energy and momentum \cite{Provost2015}. Whence the interest in examining the uniqueness of the latter kind of equations. 

\bigskip 
\appendix
\section{Appendix: Regular domains}\label{Domains}
\paragraph{Definition 1 \cite{BergerGostiaux}.}\label{Definition 1} {\it Let \Couleur{$\mathrm{M}$} be a differentiable manifold, with dimension \Couleur{$d$}. One will call} closed domain {\it of \Couleur{$\mathrm{M}$}, any closed subset \Couleur{$\mathrm{D}$} of \Couleur{$\mathrm{M}$} such that, for any \Couleur{$x\in \mathrm{D}$}, 

either} (i) {\it there is an open subset \Couleur{$\mathrm{W}$} of \Couleur{$\mathrm{M}$} such that \Couleur{$x\in \mathrm{W}\subset \mathrm{D}$},

or }(ii) {\it there is a chart \Couleur{$(\mathrm{W},\varphi )$} with \Couleur{$\varphi (x)={\bf 0}$} and \Couleur{$\varphi (y)=(y^1,...,y^d)$} for \Couleur{$y\in \mathrm{W}$}, such that}
\be\label{Case ii}
\Couleur{\mathrm{W}\cap \mathrm{D}= \{y\in \mathrm{W} ; \ y^1\leq 0 \}},\quad \mathrm{i.e.} \quad \Couleur{\varphi (\mathrm{W}\cap \mathrm{D})= \{{\bf y}\in \varphi (\mathrm{W}) ; \ y^1\leq 0 \}}.
\ee

\paragraph{Proposition 1 \cite{BergerGostiaux}.}\label{Proposition 1} {\it Let \Couleur{$\mathrm{D}$} be a closed domain of a \Couleur{$d$}--dimensional differentiable manifold \Couleur{$\mathrm{M}$}. In case} (i), {\it the point \Couleur{$x$} is in \Couleur{$\interior{\mathrm{D}}$}, the interior of \Couleur{$\mathrm{D}$} (i.e. the largest open set \Couleur{$\mathrm{U}$} of \Couleur{$\mathrm{M}$}, such that \Couleur{$\mathrm{U}\subset \mathrm{D}$}). In case } (ii), {\it the point \Couleur{$x$} is in \Couleur{$\partial \mathrm{D}$}, the boundary of \Couleur{$\mathrm{D}$}, which is a \Couleur{$(d-1)$}--dimensional submanifold of the differentiable manifold \Couleur{$\mathrm{M}$}.}\\

\noi Recall that the (topological) boundary of any subset \Couleur{$\mathrm{A}$} of \Couleur{$\mathrm{M}$} is defined to be \Couleur{$\partial \mathrm{A}:=\overline{\mathrm{A}}\cap \overline{\complement \mathrm{A}}$}, where the overbar means the adherence (or closure) in \Couleur{$\mathrm{M}$} and \Couleur{$\complement \mathrm{A}$} means the complementary set of \Couleur{$\mathrm{A}$} in \Couleur{$\mathrm{M}$}. It is easy to prove (cf. \cite{DieudonneTome1}) that we have always 
\be\label{int A inter dA = empty}
\Couleur{\interior{\mathrm{A}}\cap  \partial \mathrm{A} = \emptyset},
\ee
\be\label{int A U dA = Abar}
\Couleur{\interior{\mathrm{A}}\cup \partial \mathrm{A} = \overline{\mathrm{A}}}.
\ee
If \Couleur{$\mathrm{M}$} is an oriented manifold, then the Stokes theorem (and thus also the divergence theorem) applies to any differential \Couleur{$(d-1)$}--form (respectively to any continuously differentiable vector field), in any closed domain \Couleur{$\mathrm{D}$} of \Couleur{$\mathrm{M}$} with its boundary \Couleur{$\partial \mathrm{D}$} \cite{BergerGostiaux}. 

\paragraph{Proposition 2.}\label{Proposition 2} {\it Let \Couleur{$\mathrm{D}$} be a} closed domain {\it (see \hyperref[Definition 1]{Definition 1}) of a \Couleur{$d$}--dimensional differentiable manifold \Couleur{$\mathrm{M}$} and let \Couleur{$\mathrm{U}:=\interior{\mathrm{D}}$} be its interior. We have 
\be\label{U bar = D}
\Couleur{\overline{\mathrm{U}}=\mathrm{D}},
\ee
i.e., a closed domain is the closure of its interior. We thus call \Couleur{$\mathrm{U}:=\interior{\mathrm{D}}$} an} open domain {\it of \Couleur{$\mathrm{M}$}. Moreover, we have}
\be\label{C D bar = C U}
\Couleur{ \overline{\complement \mathrm{D}}= \complement \mathrm{U} }.
\ee

\noi {\it Proof.} Since \Couleur{$\mathrm{D}$} is a closed set such that \Couleur{$\mathrm{U}\subset \mathrm{D}$}, we have \Couleur{$\overline{\mathrm{U}}\subset \mathrm{D}$}. Due to (\ref{int A U dA = Abar}), in order to prove that \Couleur{$\mathrm{D} \subset \overline{\mathrm{U}}$}, we just have to prove that \Couleur{$ \partial \mathrm{D}\subset \overline{\mathrm{U}}$}. If \Couleur{$x\in \partial \mathrm{D}$}, we may apply to it Case (ii) of \hyperref[Definition 1]{Definition 1}. Let \Couleur{$ \mathrm{A}$} be any open neighborhood of \Couleur{$x$}. We will show that it intersects both \Couleur{$\mathrm{U}$} and \Couleur{$\complement \mathrm{D}$}; to show this, we may assume that \Couleur{$ \mathrm{A}\subset \mathrm{W}$}, with  \Couleur{$ \mathrm{W}$} the domain of the chart  \Couleur{$ \varphi $}. 
Thus \Couleur{$ \varphi (\mathrm{A})$} is an open neighborhood of \Couleur{$ \varphi (x)={\bf 0}$} in \Couleur{$ \mathbb{R}^d$}, hence it contains a ball \Couleur{$ \abs{y^j}<r\ (j=1,...,d)$}. {\it a}) Take first \Couleur{${\bf y}_0=(y^j)$} such that \Couleur{$-r<y^1<0$} and \Couleur{$\ y^j=0\ \mathrm{for\ } j=2,...,d$}: 
then \Couleur{${\bf y}_0\in \varphi (\mathrm{A})\subset \varphi (\mathrm{W})$}. 
But we get from (\ref{Case ii}) that any point \Couleur{${\bf y}\in \varphi (\mathrm{W})$} such that \Couleur{$y^1<0$} is in the interior of \Couleur{$\varphi (\mathrm{D}\cap  \mathrm{W})$}. Thus \Couleur{$ {\bf y}_0 \in \varphi (\mathrm{A})$} is in the interior of \Couleur{$\varphi (\mathrm{D}\cap  \mathrm{W})$}, or equivalently \Couleur{$y_0=\varphi ^{-1}({\bf y}_0)\in \mathrm{A}$} is in the interior of \Couleur{$\mathrm{D}\cap  \mathrm{W}$}, hence in \Couleur{$\mathrm{U}=\interior{\mathrm{D}}$}. So \Couleur{$ \partial \mathrm{D}\subset \overline{\mathrm{U}}$}, hence (\ref{U bar = D}) is proved. {\it b}) On the other hand, take now \Couleur{${\bf y}_1=(y^j)$} such that \Couleur{$0<y^1<r$} and \Couleur{$\ y^j=0\ \mathrm{for\ } j=2,...,d$}: also \Couleur{${\bf y}_1\in \varphi (\mathrm{A})\subset \varphi (\mathrm{W})$}, but we get from (\ref{Case ii}) that \Couleur{$y_1=\varphi ^{-1}({\bf y}_1)\in \mathrm{A}$} is in \Couleur{$\complement \mathrm{D}$}. Thus we have also \Couleur{$ \partial \mathrm{D}\subset \overline{\complement \mathrm{D}}$}, whence \Couleur{$ \overline{\complement \mathrm{D}}\cup  \partial \mathrm{D}=\overline{\complement \mathrm{D}}$}. But, from (\ref{int A inter dA = empty}) and (\ref{int A U dA = Abar}), we have \Couleur{$ \mathrm{D}\cap \complement \partial \mathrm{D}=\mathrm{U}$}, or \Couleur{$ (\complement \mathrm{D})\cup  \partial \mathrm{D}=\complement \mathrm{U}$}. Therefore,
\be
\Couleur{\complement \mathrm{U} =\overline{\complement \mathrm{U}}=\overline{\complement \mathrm{D}}\cup  \partial \mathrm{D}=\overline{\complement \mathrm{D}}},
\ee
which proves (\ref{C D bar = C U}). \hfill $\square$ 

\paragraph{Corollary.}\label{Corollary} {\it Let \Couleur{$\mathrm{U}$} be an open domain of \Couleur{$\mathrm{M}$} and let \Couleur{$f$} be a continuous real function defined in a neighborhood of \Couleur{$\mathrm{D}=\overline{\mathrm{U}}$}. In order that \Couleur{$\mathrm{Supp}\,f\subset \mathrm{D}$}, it is necessary and sufficient that \Couleur{$f(x)=0$} if \Couleur{$x\notin \mathrm{U}$}.}\\

\noi {\it Proof.} The support of \Couleur{$f$}, \Couleur{$\mathrm{Supp}\,f$}, is defined to be the smallest closed set containing the set of the points \Couleur{$x$} such that \Couleur{$f(x)\ne 0$}, or equivalently \Couleur{$\complement \mathrm{Supp}\,f$} is the largest open set \Couleur{$\Omega $} such that \Couleur{$f_{\mid \Omega }=0$}. Therefore, \Couleur{$\complement \mathrm{D}$} being an open set,
\be
\Couleur{\left(\mathrm{Supp}\,f\subset \mathrm{D}\right ) \Leftrightarrow \left (f_{\mid \complement \mathrm{D} }=0 \right )}.
\ee
Since \Couleur{$f$} is continuous, \Couleur{$f_{\mid \complement \mathrm{D}}=0$} is equivalent to \Couleur{$f_{\mid \overline{\complement \mathrm{D}}}=0$} --- that is, from (\ref{C D bar = C U}), to \Couleur{$f_{\mid \complement \mathrm{U}}=0$}. \hfill $\square$ \\

\noi {\bf Acknowledgement.} I am grateful to T. Padmanabhan for pointing out Ref. \cite{Padmanabhan2015} to me.\\

The author declares that there is no conflict of interest regarding the publication of this article.


\end{document}